# Colossal Cryogenic Electro-Optic Response Through Metastability in Strained BaTiO₃ Thin Films

*Albert Suceava, Sankalpa Hazra, Aiden Ross, Ian Reed Philippi, Dylan Sotir, Brynn Brower, Lei Ding, Yingxin Zhu, Zhiyu Zhang, Himirkanti Sarkar, Saugata Sarker, Yang Yang, Suchismita Sarker, Vladimir A. Stoica, Darrell G. Schlom, Long-Qing Chen, and Venkatraman Gopalan\**

The search for thin film electro-optic materials that can retain superior performance under cryogenic conditions has become critical for quantum computing. Barium titanate thin films show large linear electro-optic coefficients in the tetragonal phase at room temperature, which is severely degraded down to ≈200 pm V$^{-1}$ in the rhombohedral phase at cryogenic temperatures. There is immense interest in manipulating these phase transformations and retaining superior electro-optic properties down to liquid helium temperature. Utilizing the thermodynamic theory of optical properties, a large low-temperature electro-optic response is designed by engineering the energetic competition between different ferroelectric phases, leading to a low-symmetry monoclinic phase with a massive electro-optic response. The existence of this phase is demonstrated in a strain-tuned BaTiO$_3$ thin film that exhibits a linear electro-optic coefficient of 2516 ± 100 pm V$^{-1}$ at 5 K, which is an order of magnitude higher than the best reported performance thus far. Importantly, the electro-optic coefficient increases by 100 × during cooling, unlike the conventional films, where it degrades. Further, at the lowest temperature, significant higher order electro-optic responses also emerge. These results represent a new framework for designing materials with property enhancements by stabilizing highly tunable metastable phases with strain.

## 1. Introduction

The electro-optic (EO) effect in lithium niobate crystals enables our internet by encoding electrical to optical signals.[1–3] The EO effect describes the change in refractive index, $n$, of a material due to an applied electric field, $E_k$, given by $\Delta \left(\frac{1}{n^2}\right)_{ij} = r_{ijk} E_k + s_{ijkl} E_k E_l$ in nonmagnetic materials, where $r_{ijk}$ is the Pockel tensor representing the linear effect, $s_{ijkl}$ is the Kerr tensor representing the quadratic effect, and the dummy subscripts indicate crystal physics axes. This allows for the phase, amplitude, or polarization of light to be modulated by a driving electrical signal. Electro-optic materials have recently re-emerged as a key technology in the field of quantum computing, where the EO effect can be leveraged to perform microwave-to-optical transduction.[4,5] Qubits based on Josephson junctions and trapped ions utilize microwave frequencies to write and read quantum states at cryogenic temperatures.[6–9] Electro-optic

A. Suceava, S. Hazra, A. Ross, H. Sarkar, S. Sarker, V. A. Stoica,
L.-Q. Chen, V. Gopalan
Department of Materials Science and Engineering
The Pennsylvania State University
University Park, PA 16802, USA
E-mail: vgopalan@psu.edu

I. R. Philippi, B. Brower, V. Gopalan
Department of Physics
The Pennsylvania State University
University Park, PA 16802, USA

D. Sotir
Platform for the Accelerated Realization, Analysis, and Discovery of Interface Materials (PARADIM)
Cornell University
Ithaca, NY 14853, USA

D. Sotir, D. G. Schlom
Department of Materials Science and Engineering
Cornell University
Ithaca, NY 14853, USA

L. Ding, Y. Zhu, Z. Zhang, Y. Yang
Department of Engineering Science and Mechanics
The Pennsylvania State University
University Park, PA 16802, USA

S. Sarker
Cornell High Energy Synchrotron Source
Cornell University
Ithaca, NY 14853, USA

**DOI:** 10.1002/adma.202507564

materials are required for the transduction of these microwave signals (operating at millikelvin temperature) to communicate with infrared light that is the standard for optical networks (operating at room temperature). Cryogenic electro-optics is also important for on-chip quantum photonic circuits, trapped ion quantum computing schemes, and developments in low-temperature science.[5,10–14] Nonetheless, a materials gap exists, namely EO materials operating at cryogenic temperatures with low thermal budget and a small footprint directly integrated on a chip and with electro-optic coefficients of $r_{eff} > 1000$ pm V$^{-1}$. No such materials currently exist. Barium titanate ($BaTiO_3$) appears promising because of its large $r_{51} \approx 1640$ pm V$^{-1}$ at room temperature; however the best literature reported value for films of $BaTiO_3$ is $r_{eff} \approx 200$ pm V$^{-1}$ at 4 K.[14,15] In these prior studies, the temperature-dependent performance of the device revealed a nearly 3 × reduction in $r_{eff}$ following cooling from room to cryogenic temperature, owing to the occurrence of multiple ferroelectric–ferroelectric phase transitions. Attempts to suppress or overcome these transitions have not yielded significant improvements thus far. With regards to its electrical properties, the microwave dielectric properties of $BaTiO_3$ ceramics prepared under various methods have been extensively characterized, with the dielectric function generally rolling off above 10 GHz with an accompanying increase in loss.[16–23] Regardless, existing examples of electro-optic modulators based on thin film $BaTiO_3$ have promisingly demonstrated a consistent response at cryogenic temperatures under modulation frequencies up to 30 GHz, leaving the low-temperature functional optical properties as a significant area for improvement.[14]

Many high-performance piezoelectric materials are discovered at morphotropic phase boundaries, namely regions where compositional tuning forces a competition between distinct structural phases, yielding an intermediate low symmetry phase with superior properties relative to the parents.[24–28] Similarly, across thermotropic phase boundaries in $BaTiO_3$ and $KNbO_3$ single crystals, low symmetry monoclinic phases are stabilized by local strain and fields generated within frustrated domain microstructures created by cycling across thermal phase transitions.[29] In particular, the optical second harmonic generation coefficients in $BaTiO_3$ single crystals were enhanced by up to 4.4 × relative to the bulk values, and by 2.3 × in $KNbO_3$ in metastable monoclinic phases stabilized by inhomogeneous strain and fields. One strategy to engineer large nonlinear optical property enhancements is thus to stabilize metastable phases, but homogeneously over the entire sample.

We demonstrate this in $BaTiO_3$ thin films, a material recognized for having one of the largest electro-optic coefficients in its tetragonal (T) phase at room temperature in its bulk form and thus being at the forefront of the search for cryogenic electro-optic materials.[15,30–32] Using phase-field simulations, we design the optimal epitaxial strain conditions to promote the competition between ferroelectric phases near liquid helium temperatures, resulting in a cryogenic monoclinic (M) phase. The films demonstrate a 10 × improvement in the electro-optic effect over the best demonstrated values in the literature at 10 K.[14] Our findings demonstrate a new paradigm for engineering electro-optic materials under cryogenic conditions.

## 2. Temperature-Strain Phase Diagram of BaTiO$_3$

Bulk $BaTiO_3$ undergoes a series of phase transitions from cubic to tetragonal at 130 °C, tetragonal to orthorhombic at 5 °C, and orthorhombic to rhombohedral at −90 °C.[33,34] **Figure 1a** depicts the predicted temperature-dependent phase diagram of $BaTiO_3$ as a function of biaxial epitaxial strain, $\varepsilon = (a_\parallel - a_o)/a_o$, where $a_o$ is the equivalent cubic lattice parameter extrapolated from the high temperature $BaTiO_3$ cubic phase (4.008 Å) and $a_\parallel$ is the in-plane lattice parameter of the biaxially strained $BaTiO_3$. Under a biaxial compressive strain, the tetragonal $c$ (space group $P4mm$) phase becomes more stable by minimizing its elastic energy as compared to the other ferroelectric phases. As the biaxial compressive strain decreases, an energetic competition between the chemical energy, which determines the intrinsic stability of the ferroelectric phases (favoring the rhombohedral phase at low temperatures), and the elastic energy (favoring the tetragonal $c$ phase) emerges. Under certain strain conditions at cryogenic temperatures, this competition results in a compromise between the two energetic components resulting in the monoclinic $a$ (M$_a$) phase (space group $Cm$), which is a bridging phase between the tetragonal and rhombohedral phase.[35] In addition to the thermodynamically stable phases, after applying an electric field and retracting we find that there exists a field-induced metastable monoclinic phase that extends beyond the equilibrium monoclinic phase boundary, forming a monoclinic $c$ phase (M$_c$) (space group $Pm$), which acts as a bridging phase between the tetragonal and orthorhombic phase, as depicted in Figure 1b.

As the temperature decreases further, the system approaches the tetragonal (T) to monoclinic (M) phase boundary, where these competing phases are nearly degenerate in energy. In this regime, a small electric field can shift the thermodynamic stability between states inducing a polarization rotation and enabling a large electro-optic response, as depicted in Figure 1c. By integrating the thermodynamic theory of optical properties in ferroelectrics with phase-field simulations, we compute the temperature-dependent electro-optic response at 1550 nm.[36] Our results predict a large enhancement of the linear electro-optic coefficient, driven by the highly susceptible ferroelectric polarization of the monoclinic phase near the phase boundary. The

Y. Yang
Department of Nuclear Engineering
The Pennsylvania State University
University Park, PA 16802, USA

Y. Yang
Materials Research Institute
The Pennsylvania State University
University Park, PA 16802, USA

D. G. Schlom
Kavli Institute at Cornell for Nanoscale Science
Ithaca, NY 14853, USA

D. G. Schlom
Leibniz-Institut für Kristallzüchtung
Max-Born-Straße 2, 12489 Berlin, Germany

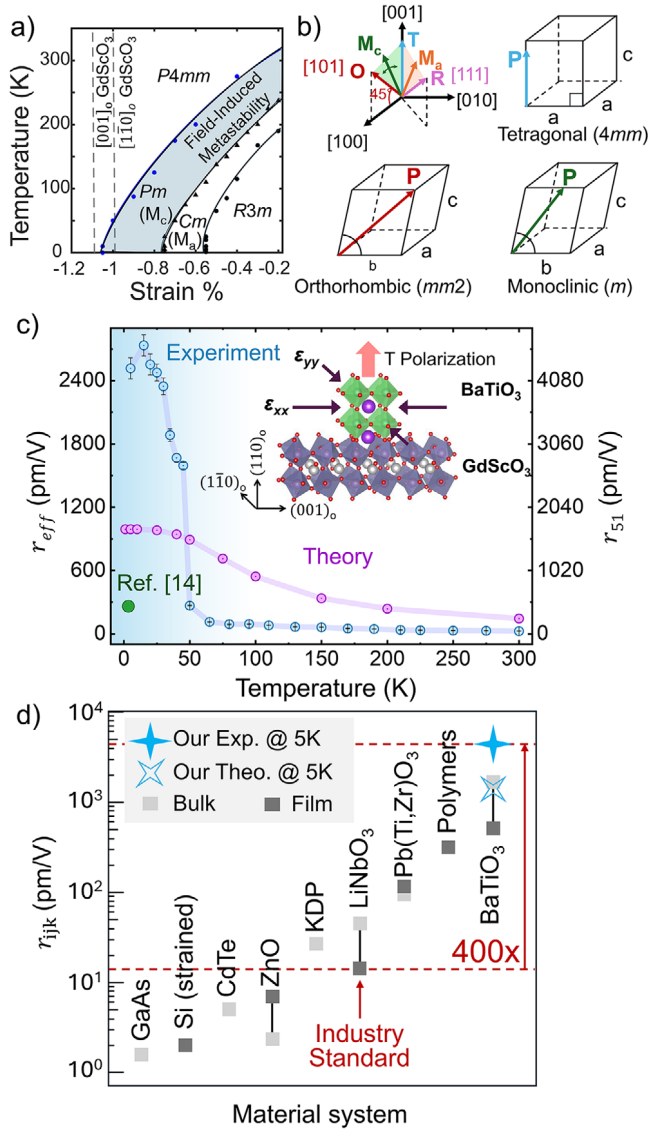

**Figure 1.** a) Temperature-strain phase diagram for compressively strained BaTiO$_3$ from phase-field simulations. The blue region indicates a metastable monoclinic phase that may be created by applying an electric field perpendicular to the tetragonal axis and removing it subsequently. b) Representation of polarization directions in tetragonal, orthorhombic, and intermediate monoclinic phases. c) Comparison between experimentally measured (blue) and phase-field simulation (purple) curves of the effective electro-optic coefficient responsible for the index change observed under the experimental geometry of Figure 2a. Comparison to the leading cryogenic coefficient found in literature is provided in (green).[14] (inset) Schematic of epitaxial BaTiO$_3$ under a compressive biaxial strain. d) Comparison between the maximum electro-optic response obtained in this work and several other benchmark materials.

general behavior may be understood using the thermodynamic theory of the electro-optic effect:

$$r_{131} = \left(\frac{\partial B_{13}}{\partial E_1}\right) \cong \left(\frac{\partial B_{13}}{\partial P_1^L}\right)\left(\frac{\partial P_1^L}{\partial E_1}\right) = f_{131}^L \chi_{11}^L \quad (1)$$

where $r_{ijk}$ is the electro-optic coefficient, $B_{ij} = \left(\frac{1}{n^2}\right)_{ij}$ is the optical dielectric stiffness defined in terms of the refractive index $n$, $P_i^L$ is the lattice polarization (the ionic and electronic components of ferroelectric polarization, which arises due to the displacement of the lattice), and $E_i$ is the electric field. We decompose the electro-optic coefficient into a linear polar-optic effect ($f_{131}^L$) and the dielectric susceptibility of the lattice polarization ($\chi_{11}^L$). The linear polar-optic effect is proportional to the spontaneous polarization and remains roughly constant down to low temperature in the compressively strained films. Therefore, the enhancement of the electro-optic effect is primarily driven by the increase of $\chi_{11}^L$ near the tetragonal-to-monoclinic phase boundary, where the lattice polarization is highly susceptible to rotation within the monoclinic mirror plane $m$.

## 3. The Electro-Optic Effect in Strained BaTiO$_3$

### 3.1. Cryogenic Electro-Optic Response

To validate this design approach, we characterize the magnitude of the Pockels effect in BaTiO$_3$ grown on (110)$_o$ GdScO$_3$ (subscript "o" for orthorhombic), where lattice mismatch results in a compressive strain of −1.0%. Epitaxial growth of a ≈37 nm film was achieved by using molecular-beam epitaxy with in-situ reflection high-energy electron diffraction (RHEED) to monitor surface quality. As shown in Figures S1 and S2 (Supporting Information), asymmetric reciprocal space maps taken at the GdScO$_3$ 332$_o$ peak reveal alignment with the BaTiO$_3$ 103$_t$ (subcript "t" for tetragonal) peak in $Q_x$, confirming epitaxial growth. $\theta$-$2\theta$ X-ray diffraction scans confirm $c$-axis out-of-plane orientation of the films in the room temperature T phase, with rocking curve measurements demonstrating film peaks with full-width half-max measures of 119 arcsec as shown in Figure S3 (Supporting Information). Post growth atomic force microscope scans reveal a root-mean-squared surface roughness of 392 pm as shown in Figure S4 (Supporting Information), confirming a smooth film morphology. Piezoelectric force microscopy scans shown in Figure S5 (Supporting Information) failed to provide evidence for antipolar domain structures. High-resolution scanning transmission electron microscopy images confirm coherent epitaxial growth and high interfacial quality, as shown in Figure S6 (Supporting Information). Additional experimental details on film growth and structural characterization are provided in Section 5.

Due to the compressive strain enforced by the substrate, at room temperature, the BaTiO$_3$ films are entirely tetragonal with the polar [001] axis in the out-of-plane direction. 500 nm square electrodes of 80 nm Pt/5 nm Ti were lithographically deposited onto the film surface with a separation gap of 200 μm to apply a field in the in-plane direction, corresponding to the [100] crystallographic direction, with a schematic of this shown in Figures S7 and S8 (Supporting Information). In the T phase of BaTiO$_3$, the crystal physics 1, 2, and 3 directions are conveniently aligned with the crystallographic [100], [010], and [001] directions.

Electro-optic measurements were performed using a polarizer-sample-compensator-analyzer (PSCA) based measurement setup, as shown in **Figure 2a**.[37–52] In this method, the change in transmission of light through a pair of crossed polarizers as a function of applied electric field is detected to quantify the change in the refractive indices of the sample. Further details on the operating principle are provided in Section 4 and within ref. [37–52]. A knife-edge characterization of the probe beam is provided in Figure S9 (Supporting Information). Other measurement methods, such as the Teng–Man approach, were considered, but the PSCA method ultimately pursued due to the restrictions the epitaxial strain condition imposes on the electrode geometry. Confidence in the measurement procedure was established by determining the electro-optic coefficient of a well-known $LiNbO_3$ reference sample, whose data are included in Figure S10 (Supporting Information).

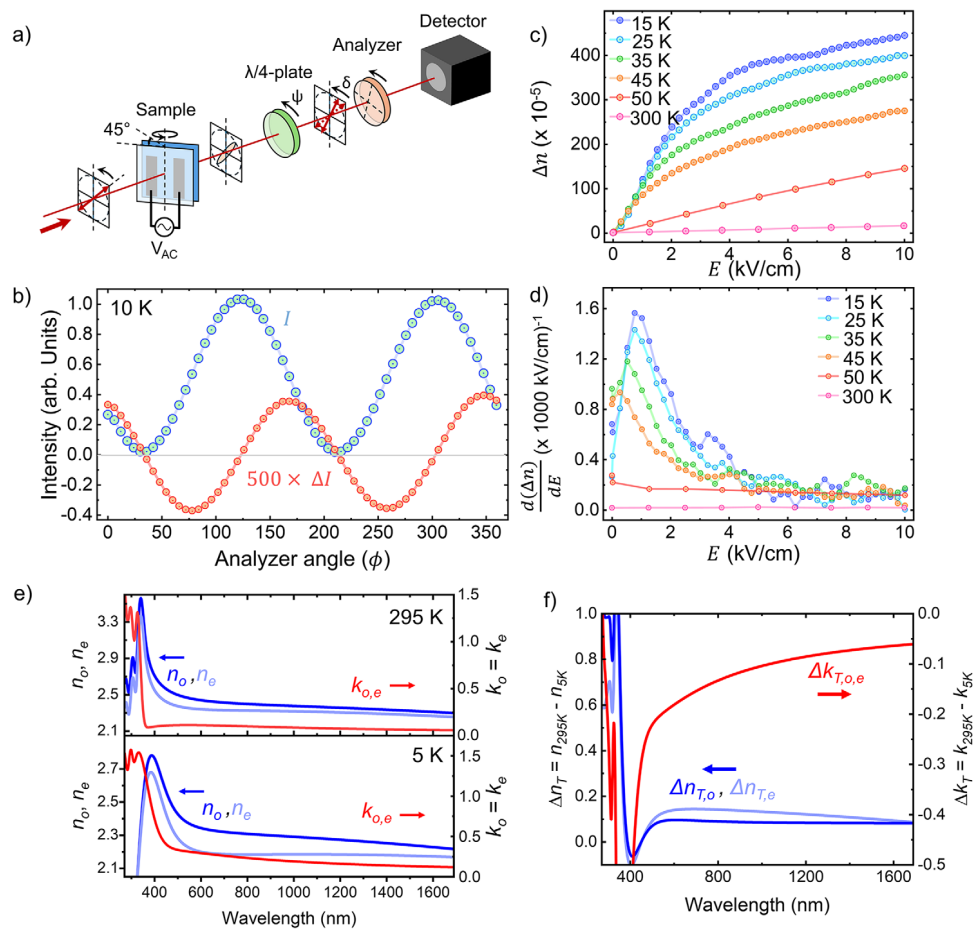

**Figure 2.** a) Schematic of experimental PCSA setup for measuring the electro-optic coefficient. b) Experimental unmodulated and modulated intensity curves, displaying the expected $\pi/2$ relative phase shift. c) $\Delta n$ versus applied electric field taken at the analyzer angle which maximizes the modulated intensity at different temperatures. d) The derivative of $\Delta n$ versus applied electric field curves, showing the offset bias required for maximum modulation efficiency. e) Refractive index and extinction coefficient of strained $BaTiO_3$ film versus wavelength, at room temperature and at 5 K, as determined by variable angle ellipsometry. f) Relative refractive index and extinction coefficient changes between room temperature and 5 K, $\Delta n_T = n_{295K} - n_{5K}$, $\Delta k = k_{295K} - k_{5K}$.

For characterization of the $BaTiO_3$ on $GdScO_3$ sample of interest, the incident polarization is set to lie between two principal optic axes of the sample, which in this study are the [100] and [010] in-plane crystallographic axes of the $BaTiO_3$ film. The $BaTiO_3$ thin film refractive indices were characterized using spectroscopic ellipsometry at room temperature and at 5 K. The extracted dispersion curves are shown in Figure 2e, and a detailed description of the optical modeling procedure and additional reflection and transmittance measurements performed is provided in Note S3, Figures S11 and S12 (Supporting Information). The ellipsometry measurements also served to refine the film thickness to a value of 36.53 nm, which was used in the subsequent analysis. With an in-plane electric field $E_1$, the index ellipsoid of the tetragonal phase distorts following Equation (2), taking the original 1, 2, and 3 principal axes to a new set of axes: $1'$, $2' = 2$, $3'$:

$$\frac{x^2}{n_1^2} + \frac{y^2}{n_2^2} + \frac{z^2}{n_3^2} + 2xz\, r_{51}\, E_1 = 1 \tag{2}$$

In the absence of the field $E_1$, $x \equiv 1$, $y \equiv 2$, $z \equiv 3$ are the principal (Eigen) crystal physics axes, and $n_1$, $n_2$, $n_3$ are the principal refractive indices along those directions with $n_1 = n_2 = n_o$ and $n_3 = n_e$ for the T phase of $BaTiO_3$, a negative uniaxial material.[1] The action of $r_{51}$ (the last term in Equation (2)) would lead to new Eigen coordinates in the x-z plane, leading to $x' \equiv 1'$ and $z' \equiv 3'$ axes. The $y' \equiv 2' = y \equiv 2$ would remain unchanged under this field. This point is illustrated in detail in Note S4 and Figure S15 (Supporting Information).

To detect the electro-optic effect under this geometry, where the $z \equiv 3$ axis is pointed normal to the film surface, the sample was tilted 45° away from the incident probe as depicted in Figure 2a. This results in the probe beam now experiencing two orthogonal refractive indices $n_o$ and $n_e(\theta)$ without any external voltage applied to the sample, and $n_o' = n_o$ and $n_e(\theta)'$ with the voltage applied. The birefringence induced by the sample results in a change in the probe polarization from a linear to an elliptical state, resulting in a non-zero transmission through an analyzer oriented orthogonally to the initial polarization state when a voltage is applied.

The unmodulated transmission function, $I$, of this system without an applied electric field is described $I = I_0 \cos^2(\phi + \delta)$, where, $\phi$ represents the angle through which the analyzer is rotated and $\delta$ represents an overall phase angle accumulated during beam transmission through the sample and other optical components before the analyzer, as shown in Figure 2a. When an applied electric field is applied across the sample, the resulting modulation to the transmission function, $\Delta I$, can be expressed with respect to the first order in $\Delta \delta$ as follows:

$$\Delta I = I_0 \, \Delta \delta \, \sin 2(\phi + \delta) \quad (3)$$

Figure 2b shows the experimentally measured $I$ and $\Delta I$ curves at 10 K for the BaTiO$_3$ on GdScO$_3$ sample fitted to the above equations. The derived $\Delta \delta$ can be converted into the electric-field induced birefringence, $\Delta n = \frac{\Delta \delta \lambda}{2\pi L}$, where, $\lambda = 1550$ nm is the wavelength of light used for measurements, and $L$ is the distance light travelled through the BaTiO$_3$ film given by $L = t/\cos \theta_{film}$, where $t$ is the film thickness and $\theta_{film}$ is the angle describing the direction of propagation of light inside the film layer, considering the 45° probe angle of incidence and Snell's law and assuming a single pass. The effective electro-optic coefficient, $r_{eff} = \frac{2\Delta n}{n_e(\theta_{film})^3 E}$ is then calculated, where $E$ is the amplitude of the AC bias applied to the sample to measure the modulated transmission function, $\Delta I$ and $n_e(\theta_{film})$ is the refractive index seen by p-polarized light, described at length in Note S4 (Supporting Information). Following Equation (3), the experimentally measured maximum value for the modulated transmission function, $\Delta I$, is obtained at the analyzer angle that results in the halfway point of the unmodulated transmission function, $I$, which is where the slope is maximized resulting in a $\frac{\pi}{2}$ phase shift between the two curves. A linear dependence in the maximum modulated intensity plotted against the applied electric field above 50 K (Figure 2c), confirms the origin of the response as the Pockels effect. The nonlinear behavior that emerges below 50 K is discussed in the following sections. This measurement is performed at several discrete temperatures to characterize the low-temperature properties of the sample, with the compensator position, $\psi$, optimized to cancel out the native birefringence of the sample at every temperature. The effective electro-optic coefficient observed, $r_{eff}$, is converted into the tensor coefficient $r_{51}$ through a geometrical analysis of the modulated index ellipsoid, as described in Note S4 (Supporting Information).

The unmodulated refractive index was directly measured at room temperature and at 5 K. For the analysis described in the preceding section, the refractive index measured at 5 K was used to analyze electro-optic data collected below 50 K, while the room temperature index was used to interpret data collected above 50 K. The separation between data sets at 50 K is chosen to reflect the phase transition that occurs at that temperature, as seen in the electro-optic measurements and second harmonic generation data to follow. An analysis of the electro-optic response using a linear interpolation of these refractive index values as opposed to a step function is presented in Figure S14 (Supporting Information). With the effective electro-optic coefficient expressed in full as $r_{eff} = \frac{2}{n_e(\theta_{film})^3 E} \left( \frac{\Delta I \cdot \lambda}{I_0 \cdot 2\pi L} \right)$, the collective error can be expressed as the sum of the errors of all terms in the products weighed by their exponents: $\frac{dr_{eff}}{r_{eff}} = 3\left(\frac{dn_o}{n_o}\right) + \frac{dE}{E} + \frac{d\Delta I}{\Delta I} + \frac{dI_0}{I_0} + \frac{dL}{L}$. In other words, every percent error in the unmodulated refractive index will propagate a three times larger percent error in the extracted electro-optic coefficient. Assuming that the refractive index in the temperature range between 5 and 295 K lies between the two endpoints measured, this results in a maximum error in the index equal to the difference between the two values, which is an error of 3.4%. Thus, the maximum error in the determination of the electro-optic coefficients due to inaccuracy in the index interpolation becomes 10.2%. An increase in the extinction coefficient of the film at low temperature may arise from the transition from a unipolar tetragonal structure to a multidomain monoclinic structure with potentially four domain variants, which might promote additional scattering of light from the domain walls.[53,54]

As shown in Figure 1c, the measured effective electro-optic coefficient reaches a peak value of 2735 ± 100 pm V$^{-1}$ at a temperature of 15 K, representing over a 100× enhancement from the room temperature value of 25 ± 2 pm V$^{-1}$. This is in direct contrast with previous results in the literature, where the electro-optic coefficients of relaxed BaTiO$_3$ films are reduced to nearly a third of their room temperature value at similar temperatures.[14] It is important to acknowledge that the room temperature electro-optic coefficient of the strained BaTiO$_3$ is lower than that of relaxed films grown on Si found in literature.[31] Phase-field simulations provide support for this trade-off being an intrinsic effect of the strained condition. There also exists the possibility of the grown film possessing oppositely oriented tetragonal $c$ domains resulting in a reduction in the observed electro-optic effect. Piezoelectric force microscopy was performed to image such potential antipolar domains, but no such domains were revealed, Figure S5 (Supporting Information). Nevertheless, the strained condition leads to superior performance at low temperatures. The reproducibility of this response is discussed in Note S5, Figure S17, and Table S3 (Supporting Information) with regards to both thermal and electrical cycling. While the sample appears to be remarkably robust and consistent with regards to electrical cycling, some variation in the nonlinear response is observed in every instance the sample is cooled. Most noteworthy is a reduction in the largest observed $r_{eff}$ after the sample had undergone several cooling-heating cycles, suggesting that thermal hysteresis and the initial configuration of the domain microstructure at low temperature carries significant implications for the nonlinear electro-optic response. The large enhancement at cryogenic temperatures arises from the emergence of the M$_c$ phase as discussed next.

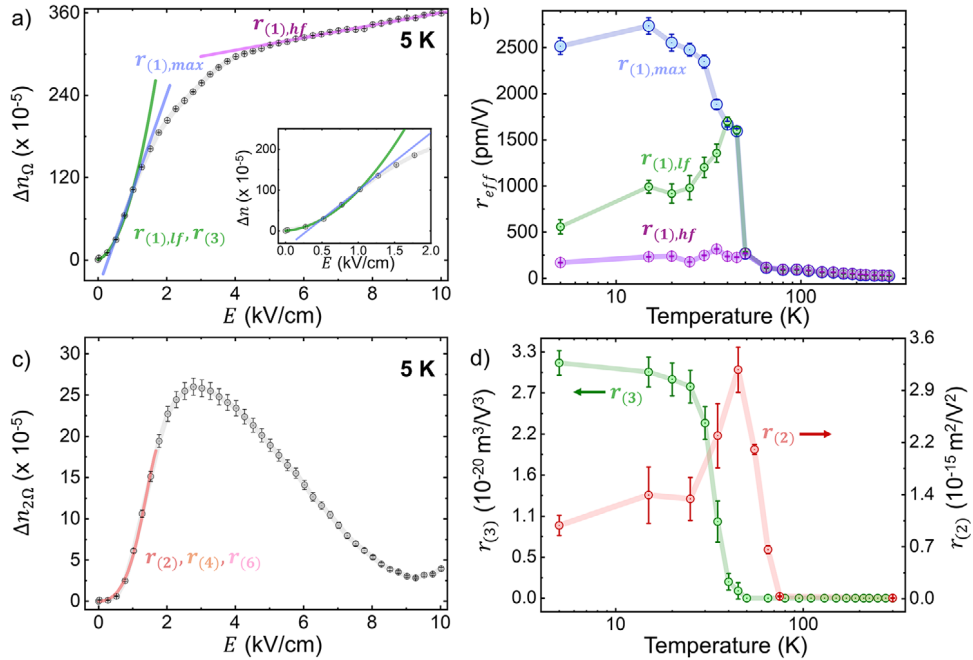

**Figure 3.** a) Various interpretations of the effective electro-optic response $r_{eff}$, depending on which part of the nonlinear response is of interest. b) $r_{(1),max}$, $r_{(1),hf}$, and $r_{(1),lf}$ as a function of temperature. c) Observed electro-optic response recorded at double the modulation frequency, $2\Omega$, as a function of temperature, indicating a nonlinear response d) Higher order electro-optic coefficients $r_{(2)}$ and $r_{(3)}$ extracted from polynomial fits of the nonlinear $2\Omega$ and $\Omega$ frequency responses, respectively.

### 3.2. Nonlinearity in the Cryogenic Electro-Optic Response

The field dependence of the electro-optic response exhibits a striking difference below 50 K from that above (Figure 2c), where a field-dependent nonlinearity is observed. Below 50 K, a knee in the curves becomes apparent, separating a linear regime at high fields from a much stronger linear response at low fields. Below 25 K, a second knee emerges at very low fields roughly below 0.5 kV cm$^{-1}$, resulting in an S-curve-like shape as shown in **Figure 3**a. Response curves of the predicted monoclinic phase can thus be discussed in terms of three different regimes: the low-field limit, where the index change appears strictly nonlinear, a mid-range high-slope linear regime containing the point of inflection for the curve, and a lower-slope high-field linear regime.

An examination of the derivative of $\Delta n$ as a function of $E$ curves (Figure 2d) reveals that the slope of the high-field regime converges to a common value, which corresponds to an $r_{eff} = r_{(1),hf} \approx 200$ pm V$^{-1}$, the temperature dependence of which is displayed in Figure 3b. The derivative of $\Delta n$ as a function of $E$ also reveals the exact point of inflection of the response curves located in the intermediate voltage range, where the magnitude of the linear electro-optic response is maximized. By taking the linear slope at the inflection point, one can define a second value for $r_{eff} = r_{(1),max}$, depicted in Figure 3a. It is this value of $r_{eff}$ that is presented in Figure 1c. From an engineering perspective, $r_{(1),max}$ can be accessed by applying a small DC bias offset alongside AC modulation to maximize the sensitivity of the electro-optic response. This DC bias would be equivalent to the point where the derivative of $\Delta n$ as a function of $E$ is maximized, as shown in Figure 2d. For response curves below 35 K, the low field regime can be fit by a higher-order polynomial function in odd powers.

The choice to only attempt a fit with odd-order polynomial terms can be understood by considering the higher-order expansion of the electro-optic effect under an AC field: $\Delta \left(\frac{1}{n(t)}\right)^2 = \sum_n r_{(n)} E(t)^n$ where $(n)$ denotes the $n^{th}$ order electro-optic effect ($r_{(n)}$ is not to be confused with $r_{ij}$ the linear electro-optic tensor coefficient). For a sinusoidal field the electro-optic response becomes: $-\Delta n(t) \frac{2}{n^3} = \sum_n r_{(n)} E_0^n \cos^n(\Omega t)$, where $\cos^n(\Omega t)$ can be expanded as:

$$\cos^n(\Omega t) = \frac{1}{2^{n-1}} \sum_{k=\frac{n}{2}+1}^{n} \binom{n}{k} \cos\left[(2k-n)\Omega t\right] + \frac{1}{2^n}\binom{n}{n/2} \text{ for even } n \quad (4)$$

$$\cos^n(\Omega t) = \frac{1}{2^{n-1}} \sum_{k=\frac{n}{2}+1}^{n} \binom{n}{k} \cos\left[(2k-n)\Omega t\right] \text{ for odd } n \quad (5)$$

It follows then that:

$$\begin{aligned}
-\Delta n(t)\frac{2}{n^3} &= \left(\tfrac{1}{2}r_{(2)}E_0^2 + \tfrac{3}{8}r_{(4)}E_0^4 + \tfrac{5}{16}r_{(6)}E_0^6 + \cdots\right) \\
&+ \left(r_{(1)}E_0 + \tfrac{3}{4}r_{(3)}E_0^3 + \tfrac{5}{8}r_{(5)}E_0^5 + \cdots\right)\cos(\Omega t) \\
&+ \left(\tfrac{1}{2}r_{(2)}E_0^2 + \tfrac{1}{2}r_{(4)}E_0^4 + \tfrac{15}{32}r_{(6)}E_0^6 + \cdots\right)\cos(2\Omega t) \quad (6)\\
&+ \left(\tfrac{1}{4}r_{(3)}E_0^3 + \tfrac{5}{16}r_{(5)}E_0^5 + \cdots\right)\cos(3\Omega t) \\
&+ \left(\tfrac{1}{8}r_{(4)}E_0^4 + \tfrac{3}{16}r_{(6)}E_0^6 + \cdots\right)\cos(4\Omega t)
\end{aligned}$$

It can be seen then that when detecting an electro-optic response at the same frequency as the driving field, only odd-powered higher order terms should manifest in the observed signal as experimentally detected by a lock-in amplifier.

A fit of up to third-order terms is sufficient for the low field regime of the electro-optic response recorded at the first harmonic of the driving electric field, with the resultant third-order electro-optic coefficients $r_{(3)}$ shown as a function of temperature in Figure 3d. Equation (6) also reveals that an intrinsically nonlinear material response is expected to yield a signal at detection frequencies that are higher harmonics of the driving electric field. Specifically, the presence of a quadratic electro-optic effect should be detected when locking on to twice the modulation frequency and integer multiples thereof. In order to confirm the nonlinear nature of the material response at cryogenic temperatures, $\Delta n$ versus $E$ curves were also measured at several discrete temperatures while locking on to double the frequency of the modulation field, $2\Omega$, with a representative result shown in Figure 3c and curves obtained at other temperatures shown in Figure S18b (Supporting Information). A fit of up to 6th order terms yields very good agreement with the data at low fields. We note that to fit the response curve for the full range of voltages applied, yet even higher order terms are required. The extracted quadratic electro-optic coefficients $r_{(2)}$ are provided in Figure 3d, while the temperature dependence of the higher order terms $r_{(4)}$ and $r_{(6)}$ is provided in Figure S18a (Supporting Information). A phenomenological model based on the Avrami equation is also capable of providing an approximate fit to the nonlinear response, as shown in Figure S19 (Supporting Information). This model is further discussed in Note S6 (Supporting Information).

Comparing different methods of defining the linear $r_{eff}$ as $r_{(1),max}$, $r_{(1),hf}$, and $r_{(1),lf}$ reveals a convergence above 50 K when the nonlinearity in the response curves vanishes. The $r_{(1),lf}$ converges to $r_{(1),max}$ with increasing temperature, while at low temperatures it approaches $r_{(1),hf}$, suggesting that as the S-curve character grows more prominent, the low-field regime may mirror the high-field regime. The relationship between the Pockels $r_{eff}$ and individual tensor elements are discussed in detail in Note S4 (Supporting Information), which includes a discussion on all possible tensor element contributions to the observed response and experiments conducted in an attempt to isolate the dominant coefficients. Following these considerations and supported by phase field simulations shown in Figure S16 (Supporting Information), the $r_{eff}$ is found to primarily reflect the intrinsic $r_{51}$ Pockels coefficient reduced by a factor of $\approx 1.7\times$, yielding a maximum value of $r_{51} = 4649 \pm 170$ pm V$^{-1}$ at 15 K, as shown in Figure 1c.

The quadratic response observed at $2\Omega$ is found to maximize $\approx 50$ K where the nonlinear response initially sets in. The fitted response is found to be within the same order of magnitude as $s_{11} - s_{12} = r_{(2),11} - r_{(2),12}$ measures of the observed quadratic electro-optic effect in bulk BaTiO$_3$ ($s_{ij}$ being the more common notation for the quadratic electro-optic effect as opposed to $r_{(2),ij}$), and several orders of magnitude higher than the response of other classical nonlinear optical materials like KH$_2$PO$_4$ (KDP).[1,55] We note that while the observed $\Delta n_{2\Omega}$ continues to increase with decreasing temperature, this does not manifest as an increase in the extracted $r_{(2)}$, but rather in the higher order $r_{(4)}$ and $r_{(6)}$ terms required to yield a good fit of the complex response (Figure S18, Supporting Information). Further exploration of the full extent of this nonlinearity with regards to even higher harmonic responses and extraction of higher order electro-optic coefficients will be investigated in subsequent works.

The nonlinearity, which is primarily responsible for the cryogenic property enhancement as defined through $r_{(1),max}$ appears to be correlated to the emergence of the metastable monoclinic structure. For applied fields below the saturation regime, a continuous monoclinic distortion can be produced through a "swaying" of the ferroelectric polarization vector away from the tetragonal out-of-plane $c$-axis, with an in-plane component developing in the direction of the applied field. This results in a large lattice dielectric susceptibility and large electro-optic response. Nonetheless, polarization rotation eventually saturates near 21° (as predicted by phase-field simulation) and the lattice susceptibility is reduced leading to the high-field regime.[56]

### 3.3. Understanding the M$_c$ Phase Through Optical Second-Harmonic Generation

To confirm the presence of a new monoclinic phase below 50 K, complementary second-harmonic generation (SHG) polarimetry measurements were performed (see Section 5: Experimental Section). The SHG process describes the generation of light at frequency $2\omega$ when light of frequency $\omega$ passes through a noncentrosymmetric medium following the third-rank tensor equation: $P_i^{2\omega} = \varepsilon_o d_{ijk} E_j^\omega E_k^\omega$, where $E_j^\omega$ and $E_k^\omega$ are optical electric fields at frequency $\omega$ and polarization directions $j$ and $k$, $P_i^{2\omega}$ is the radiating nonlinear polarization generated in the material at frequency $2\omega$ and polarization direction $i$, with efficiency described by the SHG tensor coefficient $d_{ijk}$ of the material, and $\epsilon_o$ the permittivity of vacuum.[2] Owing to its nature as a third-rank tensor property similar to the Pockels effect, polar phases and transitions can be observed with extreme sensitivity.[57] At normal incidence, no SHG was observed at room temperature as expected for the tetragonal $4mm$ phase. The observation of a non-zero SHG signal in this geometry would explicitly confirm a reduction in symmetry of the film structure. Indeed, the normal incidence SHG (**Figure** 4b) is zero until 50 K, confirming a transition to a lower symmetry structure. This is consistent with the prediction of the in-plane ferroelectric polarization component, $P_x$, as a function of temperature (Figure 4a). SHG polarimetry performed at $\theta = 45°$ incidence (Figure 4c) and its modeling (see Note S7 and Table S4, Supporting Information) indicate a tetragonal $4mm$ phase until 50 K followed by a monoclinic $m$ structure with four domain variants arising from the positive and negative shear in each of the two mirror planes perpendicular to $x$ and to $y$.[58] To further support the interpretation of the $r_{(1)max}$ response as due to a monoclinic distortion under an external electric field, the electric-field-dependent SHG response is also measured and shown in Figure S20b–e (Supporting Information). Under an external field consistent with that applied during electro-optic experiments, the polarimetry curves continue to evolve in a way that requires a monoclinic $m$ model to fit, suggesting that additional monoclinic distortions occur in the softened, low-temperature lattice. Hysteresis in the samples was observed by performing both electro-optic measurements under a DC bias and SHG measurements under DC bias. Both yield minimal discernible hysteretic behavior, with the results shown in

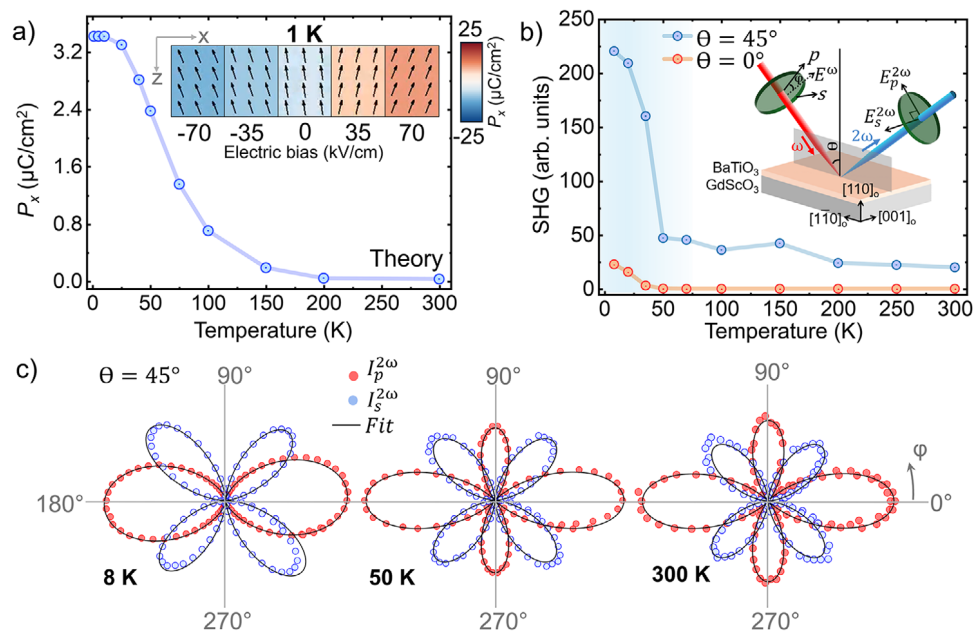

**Figure 4.** a) Phase-field simulated values of the in-plane polarization, $P_x$, as a function of temperature, reflecting the degree of monoclinic distortion. (inset) Phase field simulated polarization vectors under applied field at 1 K, showcasing the susceptibility of the lattice. b) SHG signal versus temperature at $\theta = 0°$ and $\theta = 45°$ incidence angle measurement conditions, for an input fundamental polarization, $\varphi$, along 0°. c) SHG polarimetry taken at $\theta = 45°$ incidence angle and at 300, 50, and 8 K, indicating tetragonal $4mm$ symmetry at and above 50 K and a monoclinic $m$ symmetry below 50 K. The solid lines are theory fits as described in Note S7 (Supporting Information).

Figure S21 (Supporting Information). These findings suggest that the metastable monoclinic phase enables a continuous, repeatable transformation of the polarization vector.

Additional low-temperature X-ray characterization could serve as explicit confirmation of the suspected cryogenic monoclinic phase and will be investigated in future work. Both SHG and X-ray characterization confirmed an analogous metastable phase in strained KNbO$_3$ thin films, thus strengthening the SHG-based claims presented for BaTiO$_3$.[58]

## 4. Conclusion

Through theory and experiments, we demonstrate that strain can be used to engineer intermediate low-symmetry phases in BaTiO$_3$ thin films, where a cryogenic metastable monoclinic phase exhibits a large linear electro-optic response, as well as a large nonlinear EO response. Due to the competition between multiple ferroelectric phases with similar thermodynamic stability, an emergent low-symmetry phase can be stabilized. This "bridging phase" manifests as a monoclinic structure, which is noteworthy for the ability of its lattice to deform in the presence of an applied electric field. The freedom for the polar axes to sway away from the tetragonal [001] $c$-axis with an in-plane component in the (001) plane results in a large dielectric susceptibility and a large effective electro-optic effect, the highest reported thus far. Future work is required to measure this EO response at GHz frequencies where it is of technological importance. Nonetheless, this work illustrates the power of symmetry breaking and stabilizing new metastable states in thin films toward achieving superior properties that are otherwise not available in the equilibrium phases.

## 5. Experimental Section

*Growth of BaTiO$_3$ on GdScO$_3$ Thin Films*: The sample was grown in a Veeco GEN10 MBE system equipped with an epiray GmbH THERMALAS laser substrate heater, a 1 kW CO$_2$ laser with a wavelength of 10.6 μm that irradiates the backside of the substrate over a circular area with a diameter of ≈14 mm. Barium (Sigma–Aldrich, 99.99% purity) was supplied using a conventional differentially pumped effusion cell, and titanium using a Veeco Ti-Ball source. One monolayer of BaO was deposited first, after which the sources were simply co-deposited. The barium flux was ≈5 × 10$^{13}$ atoms (cm$^2$ × s)$^{-1}$ and the titanium flux was ≈1 × 10$^{13}$ atoms (cm$^2$ × s)$^{-1}$, i.e., a Ba:Ti flux ratio of 5:1. The barium flux was determined by a quartz crystal microbalance, and the titanium flux by X-ray reflectivity of a calibration film of BaTiO$_3$ grown on an SrTiO$_3$ (001) substrate at $T_{sub}$ = 1200 °C, with an oxidant pressure of 1 × 10$^{-6}$ Torr of O$_2$ + 10% O$_3$ and a Ba:Ti ratio of 5:1. The BaTiO$_3$ film was grown on a GdScO$_3$ (110)$_o$ substrate (CrysTec GmbH), to a thickness of ≈36 nm.

Ozone was used as the oxidant at a background pressure of 1×10$^{-6}$ Torr of O$_2$ + 10% O$_3$. The film was cooled in the same oxidant and pressure in which it was grown to $T_{sub}$ < 200 °C before the oxidant was turned off. The substrate temperature, measured by a pyrometer operating at 7.5 μm on the backside of the substrate during growth, was 1160 °C. Due to the high substrate temperature, the vapor pressure of barium-containing species over BaTiO$_3$, chiefly BaO, was significantly higher than that of titanium-containing species over BaTiO$_3$, so excess barium will desorb from the surface leaving behind a single-phase BaTiO$_3$ film within an adsorption-controlled growth window. As GdScO$_3$ absorbs well at 10.6 μm, the backside of the substrate was not coated.

X-ray diffraction (XRD) and reciprocal space mapping (RSM) analysis were done with a Panalytical Empyrean X-ray diffractometer using Cu $K_{\alpha 1}$ radiation. Atomic force microscopy (AFM) was performed using an Asylum Cypher ES Environmental AFM. RHEED images were taken during growth using a Staib electron source operating at 14 kV.

*High Resolution Scanning Transmission Electron Microscopy*: A sample of the BaTiO$_3$ on GdScO$_3$ film was prepared for high-resolution scanning transmission electron microscopy (HR-STEM) using focused ion beam (FIB) lift-out, followed by sequential milling at 30, 16, 5 kV, and a final cleaning at 2 kV to progressively reduce surface damage. Scanning transmission electron microscopy (STEM) and energy-dispersive X-ray spectroscopy (EDX) analysis were performed using a ThermoFisher Scientific Titan TEM operated at 300 kV, spot size of 6, camera length of 115 mm, C2 aperture of 70 μm, convergence angle of 25.2 mrad. HR-STEM was collected with a beam current of 0.07 nA and improved by performing drift correction on sequentially acquired STEM images, followed by image averaging. The resulting images were denoised using a LASSO-like soft-thresholding filter to suppress high-frequency noise while preserving structural features. The EDX results show the mass percent distribution maps of individual elements.

*Polarization-Based Electro-Optic Measurement*: Electro-optic measurements were performed using a homebuilt PSCA setup with a 1550 nm HP 81689A continuous-wave laser source and a Nirvana 2017 balanced detector. For absolute intensity measurements, a mechanical chopper was placed in the laser path and modulated at 1.54 kHz for detection with a Stanford SR830 lock-in amplifier. For measuring the modulated transmission intensity, an Agilent 33220A signal generator amplified by a Trek 610C high voltage amplifier was used to apply a 555 Hz sine wave to the sample, with the lock-in amplifier reference signal changed to the function generator reference. The nonlinear electro-optic response was verified by also collecting electro-optic modulation signal at double the frequency of the driving voltage, i.e., at 1110 Hz. Samples were cooled to liquid He temperatures using a Janis 30 continuous flow cryostat system.

The polarization of the probe was set to lie between the ordinary and extraordinary refractive indices, as depicted in Figure 2a. An analyzer placed at the end of the setup was oriented orthogonal to the initial probe polarization state, resulting in no transmission of light through the setup when the sample and compensator were not included. With the sample introduced, the transmission becomes nonzero on account of the birefringence-induced ellipticity. A quarter-wave compensator was then added immediately after the sample to cancel out the native material birefringence with no voltage applied. This returns the transmission of the setup to zero until a voltage is applied to modulate the sample refractive indices and reintroduce ellipticity to the probe.

*Second-Harmonic Generation Polarimetry*: Second-harmonic generation polarimetry experiments were performed using a fundamental wavelength of 800 nm generated by a Spectra-Physics Solstice Ace amplified Ti:Sapphire laser with a 1 kHz repetition rate and 100 fs pulse width. Light was focused onto the sample using a 10 cm focal length lens, producing a 50 μm diameter spot size as determined by knife-edge measurements. Samples were cooled to liquid He temperatures using a Janis 30 continuous flow cryostat system. A bare $GdScO_3$ substrate of the same type produces no appreciable SHG signal under the same incident power, and thus the SHG signal measured was interpreted as originating entirely from the strained $BaTiO_3$ film.

*Phase-Field Simulations*: Based on the thermodynamic theory of optical properties, a phase-field model might formulated, where the thermodynamics of the ferroelectric system was described by its free energy functional,

$$F = \int \left[ f^L(T, P_i^L) + f^e(T, P_i^e) + f^{L-e}(P_i^L, P_i^e) + f^{grad}(\nabla_i P_j^L) \right.$$
$$\left. + f^{elas}(P_i^L, P_i^e, \varepsilon_{ij}) + f^{elec}(P_i^L, P_i^e, E_i) \right] dx^3 \quad (7)$$

where $P_i^L(x_i, t)$ is the lattice polarization, $P_i^e(x_i, t)$ is the electrical polarization, and $\sigma_{ij}(x,t)$ is the stress.[36]

The evolution of the lattice polarization and electronic polarization is solved separately. The lattice polarization is computed following the relaxational approximation, where relaxational kinetics for the lattice polarization was assumed, and the mechanical displacement, electrical potential, and electrical polarization instantaneously reach equilibrium, leading to the time-dependent Ginzburg–Landau equation

$$\gamma_{ij}^L \frac{\partial P_j^L}{\partial t} = -\frac{\delta F}{\delta P_i^L} \quad (8)$$

where $-\frac{\delta F}{\delta P_i^L}$ is the thermodynamic driving force for the temporal evolution of the lattice polarization. The local stress and lattice polarization distribution was used as inputs to the electronic polarization dynamic equation

$$\mu_{ij}^e \frac{\partial^2 P_j^e}{\partial t^2} + \gamma_{ij}^e \frac{\partial P_j^e}{\partial t} = -\frac{\delta F}{\delta P_i^e} \quad (9)$$

where $-\frac{\delta F}{\delta P_i^e}$ is the thermodynamic driving force for the temporal evolution of the electronic polarization. To compute the local electronic dielectric susceptibility, Equation (7) was solved assuming a small periodic electric field is applied, which yields an analytical solution

$$\tilde{\chi}_{ij}^{e,1}(x_i, \omega) = \left[ B_{ij}^e(x_i) - \varepsilon_0 \left( i\omega \gamma_{ij}^e + \omega^2 \mu_{ij}^e \right) \right]^{-1} \quad (10)$$

which is related to the curvature of the free-energy landscape with respect to the electronic polarization by, $B_{ij}^e(x_i) = \varepsilon_0 \left( \frac{\partial^2 f}{\partial P_i^e \partial P_j^e} \right)$. Solving Equation (14) (assuming $\lambda = 1550$ nm) the local refractive indices might be computed, from the local lattice polarization and stress distribution, thereby naturally including the electro-optic effect and potential piezo-optic effects. As the lattice polarization distribution evolves in response to an applied electric field, the corresponding change to the electronic dielectric susceptibility and the refractive index was computed.

Here $f^L(P_i^L)$ describes the intrinsic stability of the lattice polarization compared to the high symmetry phase ($m\bar{3}m$) as a Taylor expansion of the polarization about the high symmetry phase, this was equivalent to the Landau free energy density:

$$f^L(T, P_i^L) = f_o + a_{ij}(T) P_i^L P_j^L + a_{ijkl} P_i^L P_j^L P_k^L P_l^L + a_{ijklmn} P_i^L P_j^L P_k^L P_l^L P_m^L P_n^L +$$
$$a_{ijklmnop} P_i^L P_j^L P_k^L P_l^L P_m^L P_n^L P_o^L P_p^L \quad (11)$$

where $a_{ij}$, $a_{ijkl}$, $a_{ijklmn}$, and $a_{ijklmnop}$ are the dielectric stiffness coefficients measured under constant stress conditions. For $BaTiO_3$, an 8th order expansion was used to describe the stability of the lattice polarization. The thermodynamic coefficients are adapted from ref. [59] to describe the effects of quantum fluctuations at cryogenic temperatures. ref. [60]

The intrinsic free energy density of the electronic polarization, in the absence of the lattice polarization, is described by

$$f^e(T, P_i^e) = \frac{1}{2\varepsilon_0} B_{ij}^{e,ref}(T) P_i^e P_j^e \quad (12)$$

where $B_{ij}^{ref}(T)$ is related to the refractive index of the equivalent high symmetry phase. The coupling energy density between the lattice and electronic polarization, which determines the electro-optic effect, is given by

$$f^{L-e}(P_i^L, P_i^e) = g_{ijkl}^{LL} P_i^L P_k^L P_i^e P_j^e \quad (13)$$

where $g_{ijkl}^{LL}$ relates the lattice polarization to the refractive index.

The gradient energy density is represented by

$$f_{grad} = \frac{1}{2} G_{ijkl} \frac{\partial P_i^L}{\partial x_j} \frac{\partial P_k^L}{\partial x_l} \quad (14)$$

where $G_{ijkl}$ is the gradient energy tensor where the non-zero coefficients are chosen to be $G_{11} = 0.6$, $G_{22} = -0.6$, and $G_{44} = 0.6$, and the units are normalized by $\alpha_1 l_o^2$ where $\alpha_1$ is the first Landau expansion coefficient and $l_o$ is chosen as 1 nm per grid.

The elastic energy is given by

$$f^{elas}\left(P_i^L, P_i^e, \sigma_{ij}\right) = C_{ijkl}\left(\epsilon_{ij} - \epsilon_{ij}^o\right) \tag{15}$$

where $C_{ijkl}$ is the elastic stiffness, $\epsilon_{ij}$ is the total strain, and $\epsilon_{ij}^o = Q_{ijkl}\sigma_{ij}P_k^L P_l^L + \frac{1}{2\epsilon_0}\pi_{ijkl}\sigma_{ij}P_k^e P_l^e$ is the eigenstrain, where $Q_{ijkl}$ is the electrostrictive coefficient and $\pi_{ijkl}$ is the piezo-optic tensor for the high-symmetry phase. The total strain was solved for a thin film boundary condition assuming that the strain relaxes to its equilibrium value at each time step; for simplicity the contribution of the electronic polarization to the eigenstrain was ignored. Further details on solving the elasticity may be found in ref. [61]

The electrostatic energy is given by

$$f^{elec} = -E_i P_i^L - E_i P_i^e - \frac{1}{2}\epsilon_0 \kappa_{ij}^b E_i E_j \tag{16}$$

where $\epsilon_0$ is the vacuum permittivity and $\kappa_{ij}^b$ is the background dielectric constant. For simplicity, to compute the evolution of the lattice polarization, only the linear contribution to the electronic polarization was included and added to the background dielectric constant yielding $f^{elec} = -E_i P_i^L - \frac{1}{2}\epsilon_0 \kappa_{ij}^b E_i E_j$, where $\kappa_{ij}^b = 10$ and is isotropic. Here $\kappa_{ij}^b$ contains contributions from the electronic contribution and the vacuum and other hard modes. Appendix A and Table S5 (Supporting Information) contain all coefficients used in the phase-field simulations.

For the simulations, a system size of $128 \Delta x_1 \times 128 \Delta x_2 \times 56 \Delta x_3$ was used. There were 12 $\Delta x_3$ grid points for the substrates where the elastic constants were assumed to be the same as the ferroelectric film, and there are 4 $\Delta x_3$ grid points acting as an air layer above the film. The thickness of BaTiO$_3$ was set to be $40\Delta x_3$. Periodic boundary conditions were used along the in-plane lateral directions, and natural boundary conditions were used along the film-substrate and film-air interfaces. The interface between the film and the substrate was assumed to be coherent, and hence the misfit strain was calculated using the equivalent cubic lattice parameters for BaTiO$_3$ and the pseudocubic lattice parameters for GdScO$_3$.[62]

$$\epsilon_{11} = \frac{a_{GdScO3}^{[110]_o} - a_{BaTiO3}^{eq}}{a_{BaTiO3}^{eq}}, \epsilon_{22} = \frac{a_{GdScO3}^{[001]_o} - a_{BaTiO3}^{eq}}{a_{BaTiO3}^{eq}}, \epsilon_{12} = \epsilon_{21} = 0 \tag{17}$$

where the lattice parameters for BaTiO$_3$ are $a_{BaTiO3}^{eq} = 4.01 \text{\AA} (1 + 1.15 \times 10^{-5}(T - 300 \text{ K}))$ and GdScO$_3$ are $a_{GdScO3}^{[110]_o} = 3.970 \text{\AA} (1 + 1.09 \times 10^{-5}(T - 300 \text{ K}))$ and $a_{GdScO3}^{[001]_o} = 3.966 \text{\AA} (1 + 1.09 \times 10^{-5}(T - 300 \text{ K}))$.

The simulations begin with an initial condition of $P_i^L(x_i, t = 0) = [0.0\ 0.0\ 0.1\ \text{C m}^{-2}] + \Delta P^{noise}(x_i, t = 0)$, where $|P^{noise}| = 0.1\ \text{C m}^{-2}$. To simulate the evolution of the polarization under an applied electric field and the corresponding electro-optic effect, a uniform electric field was applied along the [100] direction ranging from $-70$ to $70$ kV cm$^{-1}$. The electric field was incremented by $0.7$ kV cm$^{-1}$ every 100 timesteps. The linear electro-optic effect and quadratic electro-optic effect were found by taking the average of the refractive index over the ferroelectric material as a function of an applied electric field and fitting to a polynomial expansion centered around zero. The simulations for the evolution of the lattice polarization were completed using the mu-Pro software.

*Synchrotron-Based Reciprocal Space Mapping*: Synchrotron X-ray diffraction experiments were performed at the ID4B (QM2) beamline at the Cornell high energy synchrotron source (CHESS). The incident X-ray energy was 20 keV. An area detector array (Pilatus 6M) was used to collect the scattering intensities in a grazing incidence reflection geometry (8° incidence angle). The sample was rotated through 360° rotations, sliced into 0.1° frames. Geometric parameters of the Pilatus 6M detector, such as detector distance, tilting, rotation, and direct beam position, were extracted using a standard CeO$_2$ powder reference.

## Supporting Information

Supporting Information is available from the Wiley Online Library or from the author.

## Acknowledgements

A.S., S.H., and A.R. contributed equally to this work. This work was primarily supported through DOE-BES, under award No. DE-SC0012375. S.H., V.A.S., and V.G. acknowledge support from DOE-BES grant DE-SC0012375 for partial electro-optic measurements, partial SHG measurements, X-ray experiments, and manuscript preparation. A.S and V.G. acknowledge support from the Center for 3D Ferroelectric Microelectronics and Manufacturing (3DFeM$_2$), an Energy Frontier Research Center funded by the U.S. Department of Energy, Office of Science, Office of Basic Energy Sciences Energy Frontier Research Centers program under Award Number DE-SC0021118, for partial electro-optic measurements, partial SHG measurements, optical characterization, and manuscript preparation. I.R.P. and B.B. acknowledge the National Science Foundation DMREF Grant No. DMR-2522897 for partial electro-optic and spectroscopic ellipsometry measurements. S.S and V.G. acknowledge support from the National Science Foundation supported Penn State MRSEC for Nanoscale Science Grant Number DMR-2011839 for UV–vis spectroscopy measurements. A.R., L.Q.C., and V.G. acknowledge support from the U.S. Department of Energy, Office of Science, Office of Basic Energy Sciences, under Contract No. DE-SC0020145 for the phase-field simulations and manuscript preparation. A.R. also acknowledges the support of the National Science Foundation Graduate Research Fellowship Program under Grant No. DGE1255832. The phase-field simulations in this work were performed using Bridges-2 at the Pittsburg Supercomputing Center through allocation MAT230041 from the ACCESS program, which is supported by National Science Foundation Grants Nos. 2138259, 2138286, 2138307, 2137603 and 2138296. The BaTiO$_3$ thin films were synthesized at the Platform for the Accelerated Realization, Analysis, and Discovery of Interface Materials (PARADIM), which is supported by the National Science Foundation (NSF) under Cooperative Agreement No. DMR-2039380. D.S. and D.G.S. acknowledge support from the NSF through PARADIM under Cooperative Agreement No. DMR-2039380. This work is based on research conducted at the Center for High-Energy X-ray Sciences (CHEXS), which is supported by the National Science Foundation (BIO, ENG and MPS Directorates) under award DMR-2342336. This research used Electron Microscopy resources of the Center for Functional Nanomaterials (CFN), which is a U.S. Department of Energy Office of Science User Facility, at Brookhaven National Laboratory under Contract No. DE-SC0012704.

## Conflict of Interest

The authors have a provisional patent filed on this work. Long-Qing Chen has a financial interest in MuPRO, LLC, a company which licenses and markets the software package used in this research.

## Data Availability Statement

The data that supports the findings of this study are available within the article. Datasets used for the generation of figures presented in the article is available at https://doi.org/10.5281/zenodo.17296490. Additional data related to the film growth and structural characterization is available at https://doi.org/10.34863/k6vg-fr77. Any additional data connected to the study are available from the corresponding author upon reasonable request.

# Supporting Information

**Colossal Cryogenic Electro-Optic Response through Metastability in Strained BaTiO$_3$ Thin Films**

*Albert Suceava\*, Sankalpa Hazra\*, Aiden Ross\*, Ian Reed Philippi, Dylan Sotir, Brynn Brower, Lei Ding, Yingxin Zhu, Zhiyu Zhang, Himirkanti Sarkar, Saugata Sarker, Yang Yang, Suchismita Sarker, Vladimir A. Stoica, Darrell G. Schlom, Long-Qing Chen and Venkatraman Gopalan*

# Note S1: Structural Characterization of BaTiO$_3$ on GdScO$_3$ Thin Films

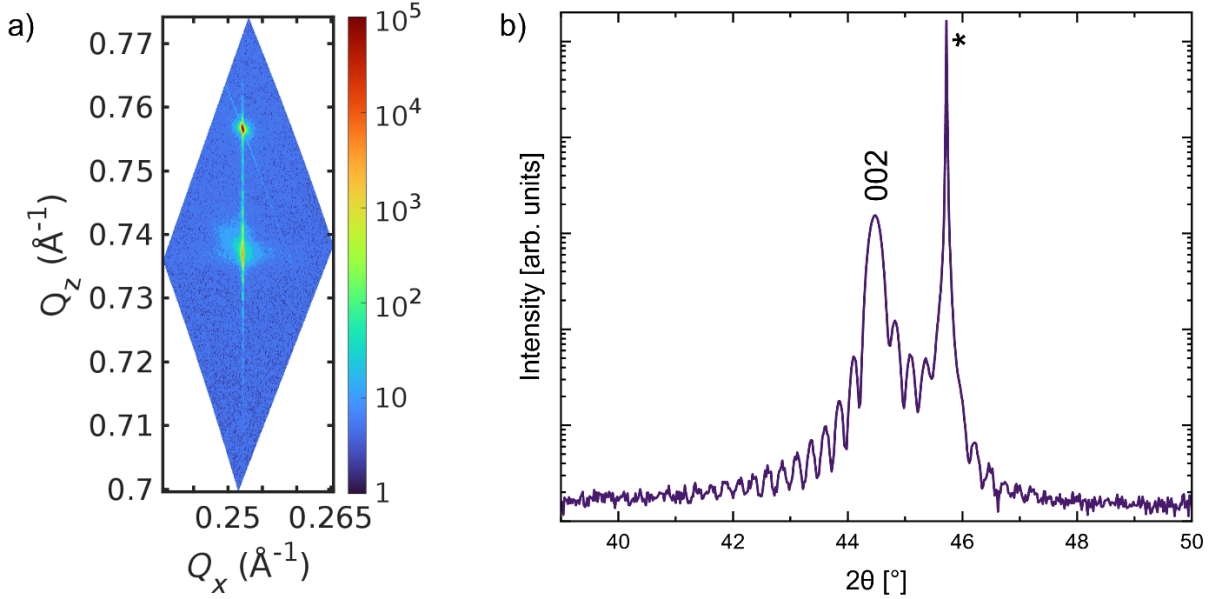

**Figure S1.** a) Asymmetric reciprocal space map scan taken at the GdScO$_3$ 332 peak, demonstrating epitaxial growth. b) $\theta$-$2\theta$ x-ray diffraction scan showing an approximately 37 nm thick film of BaTiO$_3$ grown on GdScO$_3$ (110)$_o$ at $T_{sub}$ = 1160 °C at a background pressure of 10% ozone of 1×10$^{-6}$ Torr. The $c$-axis lattice constant, measured from the RSM, is 4.073 ± 0.003 Å. The film is commensurately strained to the substrate and shows only $c$-axis orientation. This can be compared to the predicted $c$-axis lattice constant of ideal BaTiO$_3$ commensurately strained to GdScO$_3$ (110), which is calculated using the elastic stiffness tensor in Voigt notation and experimentally determined lattice parameters of BaTiO$_3$ and the GdScO$_3$ substrate (which has a congruently melting composition that differs from the stoichiometric composition):[1–3]

$$c_\perp = c_{BaTiO_3} + \frac{\left(4 a_{BaTiO_3} - c_{GdScO_3} - \sqrt{a_{GdScO_3}^2 + b_{GdScO_3}^2}\right) c_{BaTiO_3} C_{13}}{2 a_{BaTiO_3} C_{33}}$$

The predicted value is 4.077 Å, which is close to the out-of-plane lattice constant of this film. This is also in agreement with Matsubara *et al.* who demonstrated commensurately strained BaTiO$_3$ films on GdScO$_3$ (110) substrates, grown in an adsorption-controlled regime via metalorganic gas-source MBE, with a $c$-axis lattice parameter of 4.074 Å.[4]

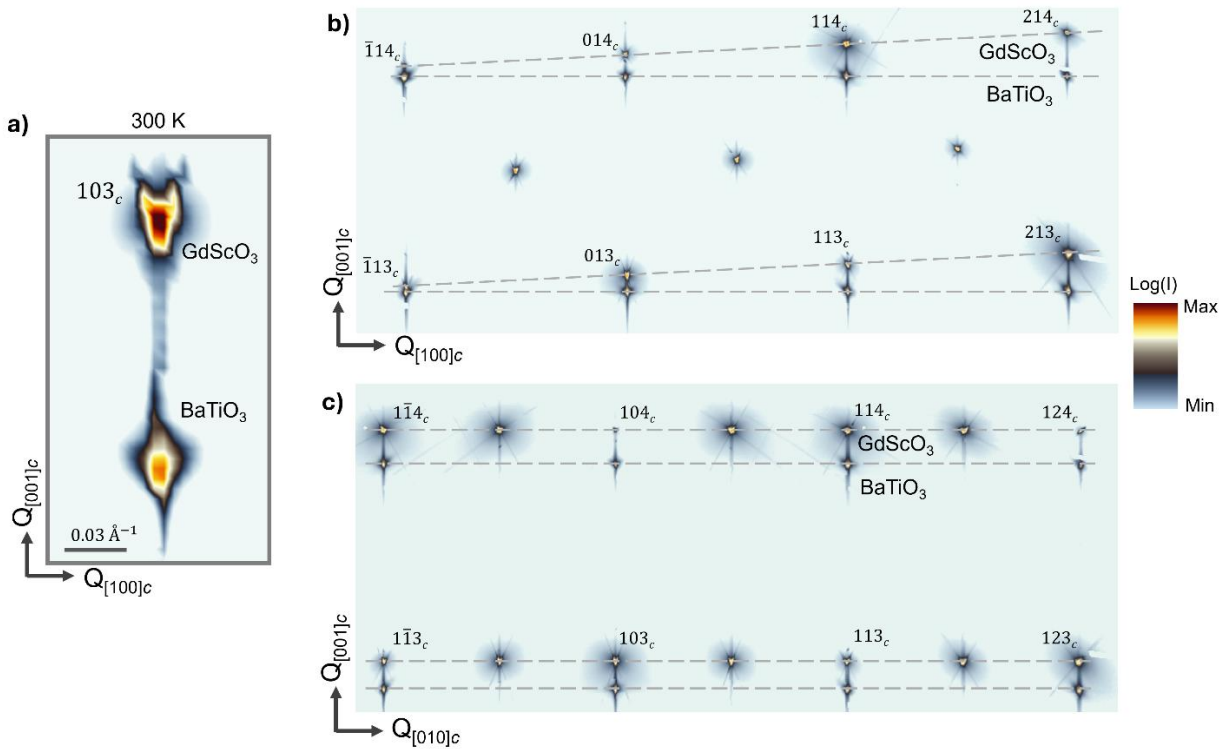

**Figure S2.** a) Room temperature X-ray reciprocal space maps (RSM) around the $103_c$ GdScO$_3$ peak (*c* stands for cubic notation) showing the epitaxially coherent nature of the BaTiO$_3$ film. b) Large area reciprocal space planes (normal to $[010]_c$ direction) showing tilted series of GdScO$_3$ peaks related to the orthorhombic nature of the substrate while the films peaks are aligned with the crystallographic direction as expected for a tetragonal structure of the BaTiO$_3$ thin films at room temperature. (c) Similar large area RSMs normal to the $[100]_c$ direction displaying no such tilts in either GdScO$_3$ substrate peaks or BaTiO$_3$ thin film peaks.

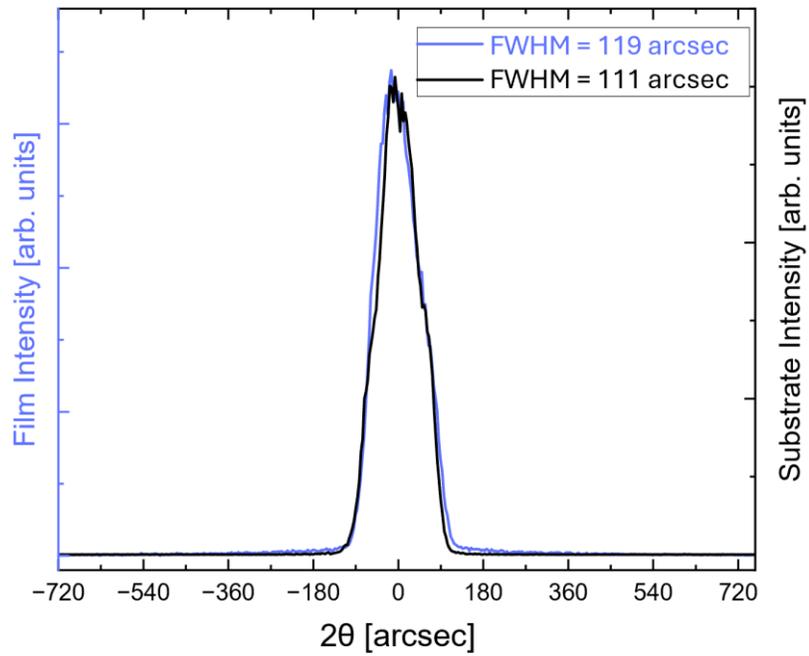

**Figure S3.** Rocking curve measurement comparing the film and the substrate, taken at the BaTiO$_3$ 002 and the GdScO$_3$ 220 diffraction peaks.

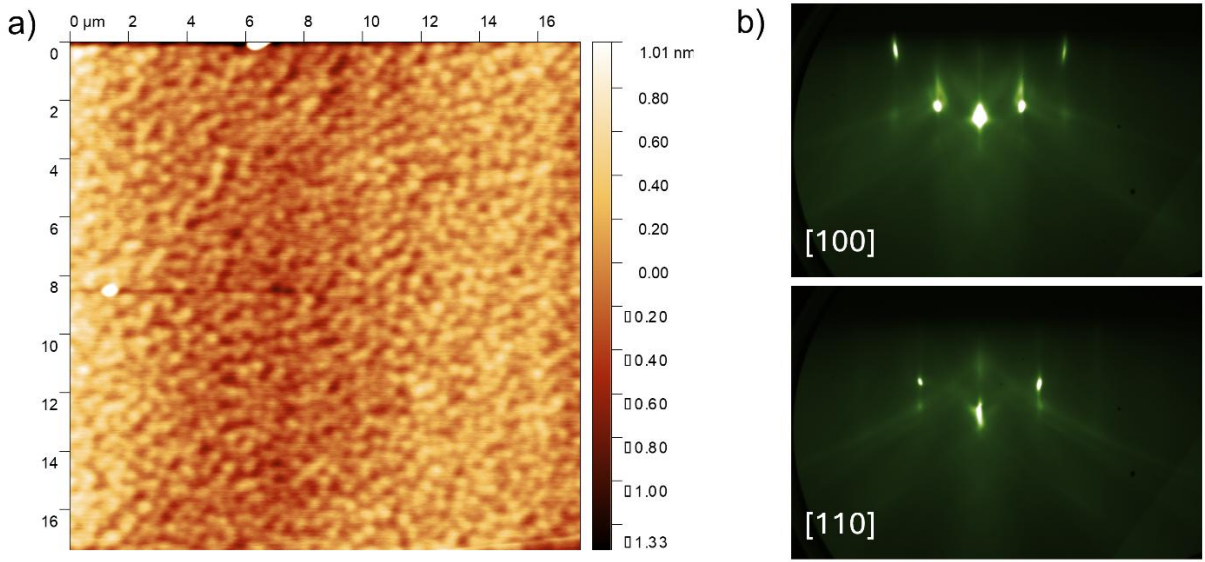

**Figure S4.** a) Atomic force microscope scan of the film surface, with a root-mean-square roughness of 392 pm. b) RHEED images of the film post growth, taken at room temperature along the azimuths of BaTiO$_3$ indicated.

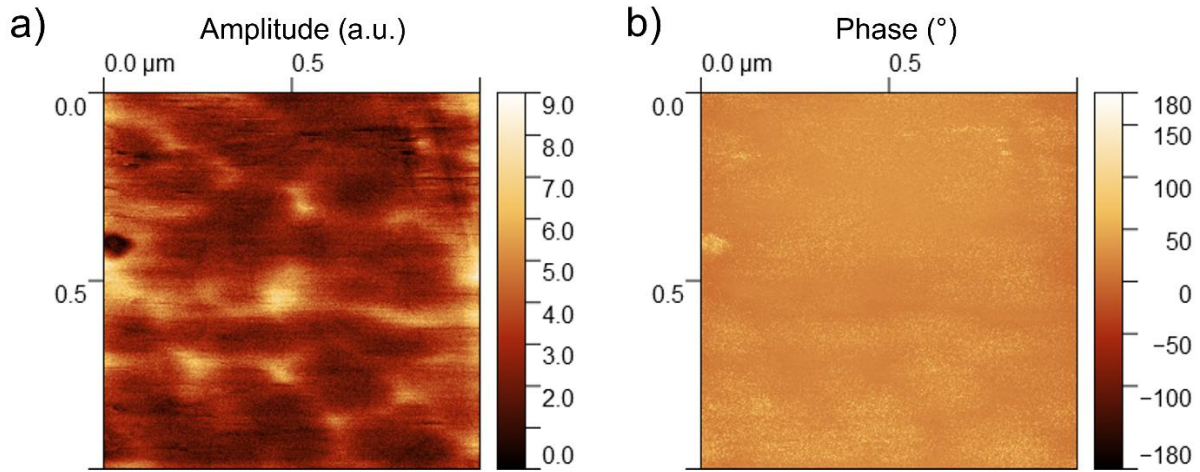

**Figure S5.** a) Amplitude and b) phase signal channels of room temperature piezoelectric force microscopy scans collected from a 1 × 1 µm region of the film. No contrast in phase is apparent, suggesting the absence of antipolar tetragonal domains.

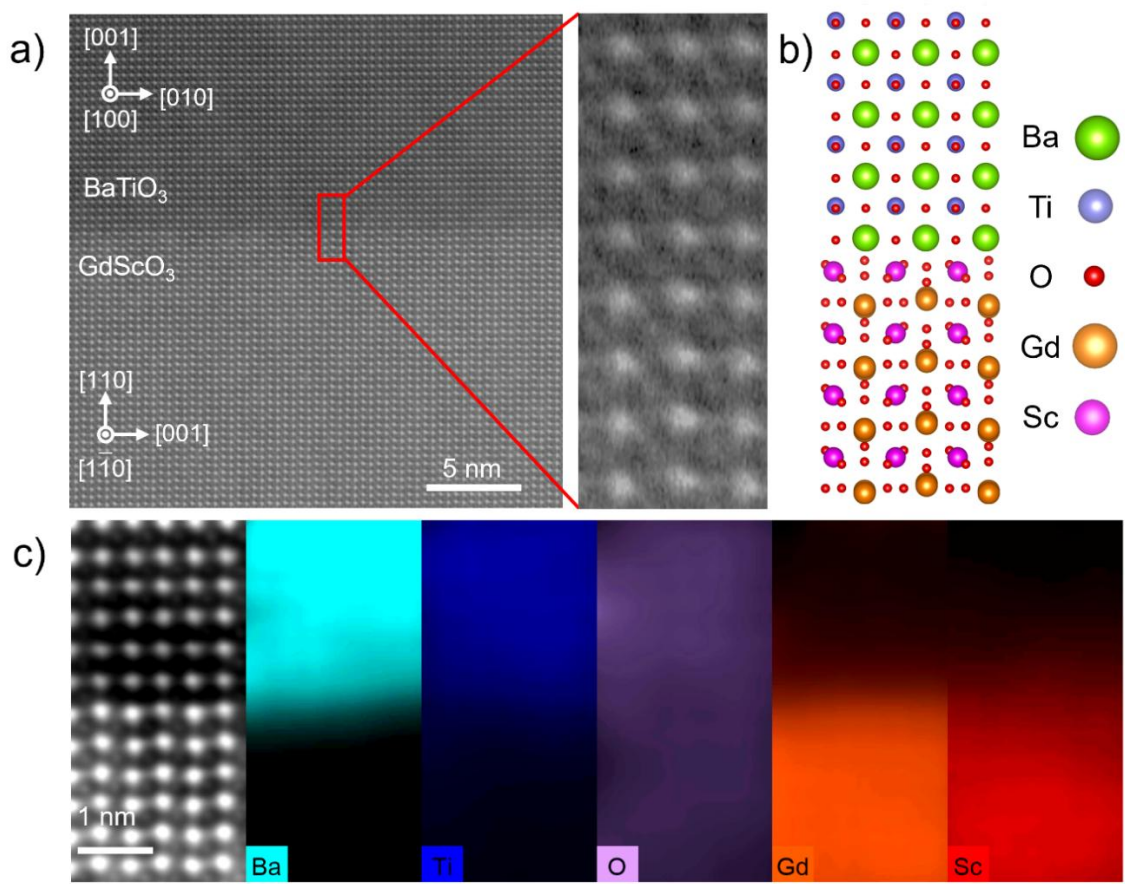

**Figure S6.** a) HRSTEM image of BaTiO$_3$ and GdScO$_3$ interface, demonstrating coherent growth. b) Schematic of interfacial structure. c) Energy-dispersive X-ray Spectroscopy images displaying the mass percent distribution maps of elemental species at the interface.

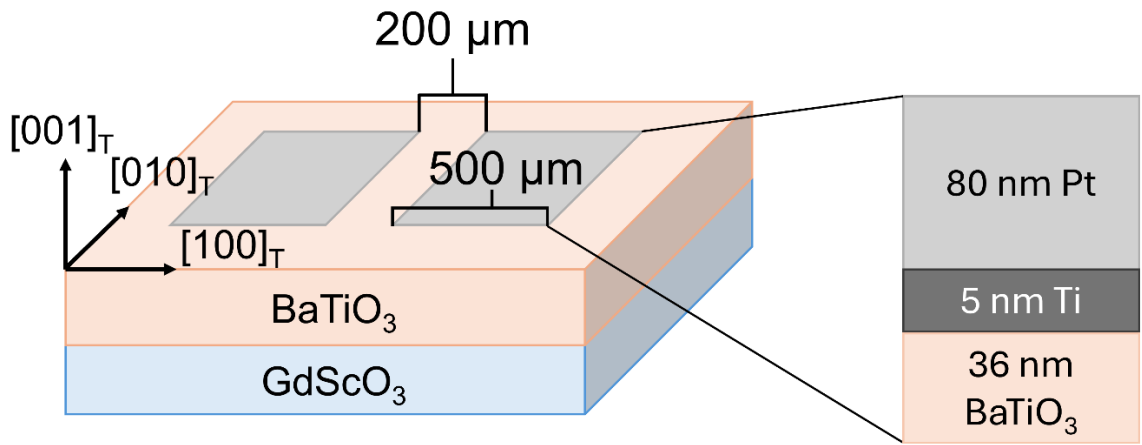

**Figure S7.** Schematic of surface deposited square electrodes for applying an in-plane electric field.

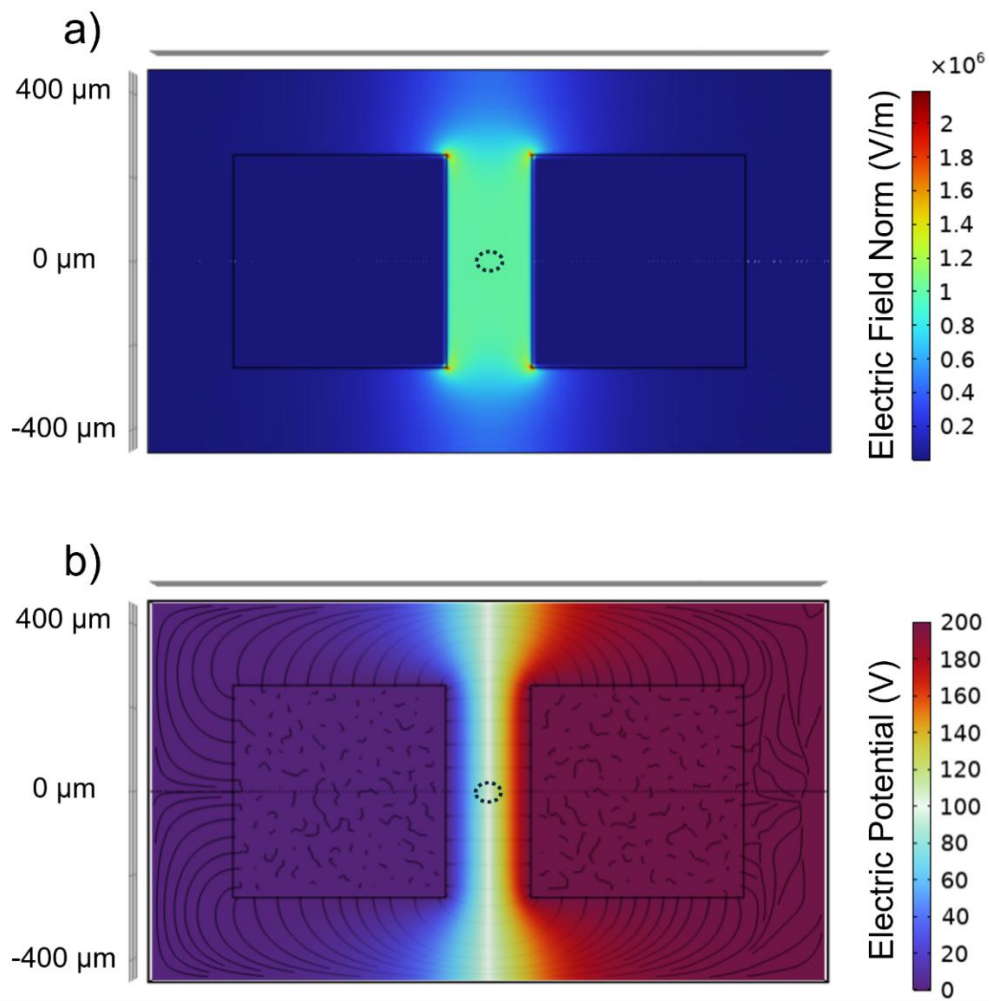

**Figure S8.** a) Electric field magnitude and (b) electric potential between surface deposited electrodes as simulated in the COMSOL Multiphysics Electrostatics module. Vector field streamlines have been added to the electric potential visualization as a guide to the eye. The simulation environment consists of a 37 nm film layer on top of a 200 nm substrate layer. Their dielectric constants are 2000 and 20 respectively [5, 6]. Two 500 μm square surfaces are defined on the film surface, with one set to ground and the other a 200 V potential. The COMSOL calculated electric field value of 10 kV/cm in the electrode gap matches the expected value obtained when dividing the voltage applied by the electrode separation. The dotted ellipses indicate the relative full width half max diameter of the focused beam used for electro-optic characterization, projected onto the film surface at 45° incidence.

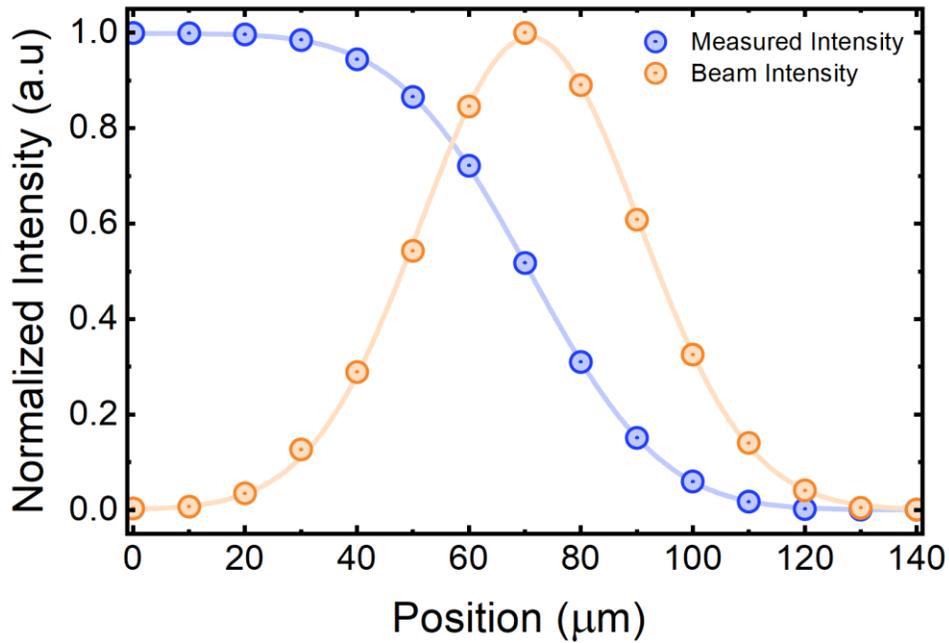

**Figure S9.** Knife edge measurements of 1550 nm probe profile utilized for electro-optic measurement. The experimentally measured signal strength is shown with blue data points, while the gradient of the experimental data reflecting the probe intensity profile is shown with orange data points. Solid lines indicate fits to data of the form $I = I_0 \cdot \frac{1}{2}\left(1 - \text{erf}\left[\frac{x-x_0}{\sigma\sqrt{2}}\right]\right)$ and $\frac{dI}{dx} = -I_0 \frac{\exp[-(x-x_0)^2/2\sigma^2]}{\sigma\sqrt{2\pi}}$ respectively. The full width half max beam diameter is 43.9 μm and the diameter measured from the $1/e^2$ level is 74.6 μm.

## Note S2: Validation of Electro-Optic Measurement Against a Standard Sample

In order to develop confidence in the function of the PSCA setup and the analysis of experimental data, measurements were performed on a reference 10x10x1 mm X-cut LiNbO$_3$ single crystal obtained from MTI Corporation. Electrodes were prepared such that a field could be applied along the [0001] crystal axis, corresponding to the crystal physics 3 direction, by blanket sputtering 100 nm of Pt on both faces of the crystal while using a strip of Kapton tape to shield a thin region in the center of each face. This preparation resulted in a 1.85 mm electrode gap and electrical continuity between sputtered Pt on both faces, minimizing fringing effects and maximizing the homogeneity of the 3-oriented field within the electrode gap.

Measurement was performed at normal incidence with an electric field applied along the crystal physics 3 direction, resulting in the birefringence of the crystal being modulated according to the $r_{13}$ and $r_{33}$ coefficients. Since $r_{13}$ and $r_{33}$ will work in unison to modulate the sample birefringence, neither coefficient can be measured in isolation; an effective response $r_{eff} = r_{33} - \left(\frac{n_o^3}{n_e^3}\right) r_{13}$ is observed. An effective electro-optic coefficient of $r_{eff} = 23.4$ pm/V can be retrieved from the slope of the linear electro-optic response shown in **Figure S9b**. This is to be compared against a predicted value of $r_{eff} = 21.3$ pm/V based on material property parameters provided by the supplier, demonstrating a reasonable level of agreement.

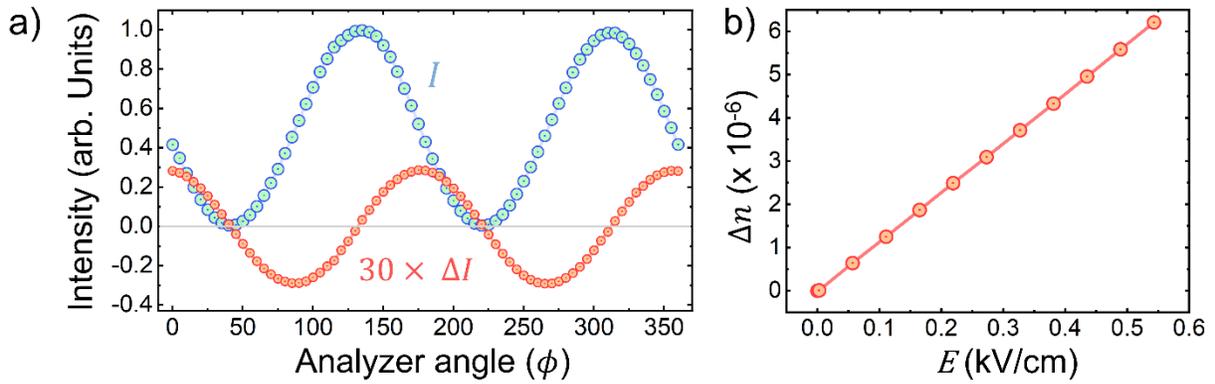

**Figure S10.** a) Modulated and unmodulated intensity curves retrieved from measurement of an X-cut LiNbO$_3$ single crystal. (b) The observed refractive index change as a function of electric field applied to the single crystal, demonstrating a clear linear response.

## Note S3: Linear Optical Properties of Strained BaTiO$_3$ Film

To quantify the room temperature refractive index of the sample, spectroscopic ellipsometry measurements were performing using a J. A. Woollam M-2000XI rotating-compensator, variable angle spectroscopic ellipsometer, with Ψ, Δ, and Mueller matrix parameters collected at incident angles of 45°, 55°, and 65°. Data was collected across the 271.1 to 1689.3 nm spectral range and fitting was performed within the J.A. Woollam CompleteEASE software. The complex refractive index values $n$ and $k$ are extracted from Ψ, Δ, and $M_{ij}$ spectra using a least-squares regression analysis and unweighted error function to fit constructed dielectric function models of a semi-infinite GdScO$_3$ substrate and BaTiO$_3$ film layers to experimental data.

An anisotropic model of the film refractive index accounting for sample birefringence was developed as follows. Firstly, measurements were performed on a sample of the GdScO$_3$ substrate and a dielectric model following the Cauchy equation: $n(\lambda) = A + \frac{B}{\lambda^2} + \frac{C}{\lambda^4}$, where $\lambda$ is in units of μm. An Urbach absorption tail was included in the model to account for a non-zero imaginary component of the refractive index, $k$, of the form: $k = k_{amp}\, e^{G(E-E_g)}$. We note that in the Urbach tail equation, $k_{amp}$ and $E_g$ are coupled variables during fitting, and so $E_g$ does not represent a physical parameter with a unique solution. Fit parameters obtained for the GdScO$_3$ substrate are: $A = 2.001, B = 0.01125, C = 0.00029005, G = 0.443, E_g = 0.443$.

Secondly, a model is constructed for fitting data collected on a sample of the BaTiO$_3$, film studied grown on GdScO$_3$, with the GdScO$_3$ dielectric function developed in previous fitting step used with the fit parameters fixed. An isotropic model was constructed from the data set collected at 45°, assuming that the sample index observed reflects $n_o$ nearly completely ($n_e$ corresponding to light polarized along the out-of-plane polar direction of the tetragonal unit cell). A dielectric function model consisting of multiple oscillator functions is then constructed. Two separate models, one consisting of Lorentz oscillators and the other consisting of Gaussian oscillators, were developed and considered. The functional form of the Lorentz oscillators is given by $\varepsilon_{Lorentz_i} = \frac{Amp_i \cdot Br_i \cdot En_i}{En_i^2 - E^2 - i \cdot E \cdot Br_i}$, with fit parameters $Amp_i$, $Br_i$, and $En_i$ representing the amplitude, width, and center position of the $i^{th}$ oscillator respectively, and $E$ the photon energy expressed in eV. The form of the Gaussian oscillators is given by $\varepsilon_{Gaussian_i} = Amp_i \left\{ \left[ \Gamma\left(\frac{E-En_i}{\sigma_i}\right) + \Gamma\left(\frac{E+En_i}{\sigma_i}\right) \right] + i \cdot \left( \exp\left[ -\left(\frac{E-En_i}{\sigma_i}\right)^2 \right] - \exp\left[ -\left(\frac{E+En_i}{\sigma_i}\right)^2 \right] \right) \right\}$, where $\sigma_i = \frac{Br_i}{2\sqrt{\ln(2)}}$ and $\Gamma$ is a convergence series that produces a Kramers-Kronig consistent line shape for $\varepsilon_1$.[7–9] The primary difference between the two oscillators lies in the peak shape in the imaginary part of the dielectric function corresponding to the resonance of the oscillator. The Gaussian oscillator provides a much steeper tail of the oscillator away from the center energy, resulting in sharper band edges and reduced extinction coefficients away from the band edge. The real part of the film dielectric function is further affected by a wavelength-independent constant, $\varepsilon_\infty$, and UV and IR poles following Lorentz oscillators with zero-broadening that reflect resonances far outside the measurement spectral range: $\varepsilon_{IR} = \frac{Amp_{IR}}{En_{IR}^2 - E^2}$ and $\varepsilon_{UV} = \frac{Amp_{UV}}{En_{UV}^2 - E^2}$. An initial guess for the film thickness of $t = 37$ nm was used, based on XRR measurements performed for initial structural characterization.

After the fit parameters $Amp_i$, $Br_i$, and $En_i$ are allowed to relax against the 45° data set, further fitting is performed against the 65° data set. The BaTiO$_3$ layer model is made anisotropic, with the extraordinary axis oriented in the out-of-plane direction and the extraordinary index determined by summing the ordinary optical constants with "difference" values calculated from an extended Cauchy dispersion equation: $d\varepsilon_i$.[7] The extended Cauchy equation resembles that used to fit the optical constants of the GdScO$_3$ substrate, with the inclusion of a higher order term and IR pole. The difference parameters serve as fitting parameters for the extraordinary index and are allowed to relax to fit the 65° incidence data set, with the ordinary optical constants fixed.

Following the initial generation of the extraordinary index, both the ordinary and extraordinary optical constants are allowed to relax to simultaneously fit all angles in the data set: 45°, 55°, and 65°. Once the optical fit parameters are relaxed against the complete data set, the optical constants are fixed and the film thickness made a new fit parameter to be relaxed until a minimum in mean-squared error is reached. Alternating relaxation of the fits for film optical constants and film thickness is performed until a global minimum in all parameters is reached. The final models developed for the film layer are described below in **Table S1** and the accompanying text:

**Table S1: Fit parameters for optical constants of BaTiO$_3$ film at room temperature**

| Oscillator Number | Lorentz | | | Gaussian | | |
|---|---|---|---|---|---|---|
| | $Amp_i$ | $Br_i$ | $En_i$ | $Amp_i$ | $Br_i$ | $En_i$ |
| 1 | 0.324128 | 5.0491 | 1.828 | 0.668082 | 5.0302 | 1.873 |
| 2 | 4.259748 | 0.1964 | 3.781 | 7.428887 | 0.2808 | 3.756 |
| 3 | 4.377518 | 0.4444 | 4.118 | 5.664778 | 0.3357 | 4.112 |
| 4 | 7.828694 | 0.6844 | 4.705 | 8.632928 | 0.6736 | 4.677 |

Additional fit parameters include:

For the Lorentz oscillator model: $t = 36.53$ nm, $\varepsilon_\infty = 0$, $Amp_{IR} = 0.2831$, $En_{IR} = 0$ eV, $Amp_{UV} = 438.5576$, $En_{UV} = 11.125$ eV, $d\varepsilon_A = -0.335015$, $d\varepsilon_B = -0.402018$, $d\varepsilon_C = 0.080858$, $d\varepsilon_D = -0.004421$, $d\varepsilon_{IR} = 0.207897$.

For the Gaussian oscillator model: $t = 36.53$ nm, $\varepsilon_\infty = 0$, $Amp_{IR} = 0.2831$, $En_{IR} = 0$ eV, $Amp_{UV} = 438.5576$, $En_{UV} = 11.125$ eV, $d\varepsilon_A = -0.029490$, $d\varepsilon_B = 0.974320$, $d\varepsilon_C = -0.060582$, $d\varepsilon_D = 0.002233$, $d\varepsilon_{IR} = -0.050425$.

For determination of the cryogenic refractive index, a modified Oxford Instruments MicrostatHe cryostat was integrated with the aforementioned Woollam M-2000XI. The cryostat windows were constructed to be normal to an incident probe at 65°, resulting only in modulation of the spectroscopic ellipsometry probe intensity without affecting measured Ψ, Δ and $M_{ij}$ quantities when data was collected at that angle. As such, Ψ, Δ and $M_{ij}$ spectra were collected for an incident angle of 65° and at 5 K. During the cooling process, the cryostat chamber was pumped to a vacuum of $10^{-5}$ at room temperature and no sample icing was observed, which would lead to periodic fluctuations in Ψ and Δ with wavelength if at least one the order of a hundred nanometers in thickness. Thus, all changes in Ψ, Δ and $M_{ij}$ measured at 5 K are attributed to changes in the refractive index of the BaTiO₃ film. The existing room temperature birefringent model was allowed to relax in order to best fit the low temperature data. While a monoclinic structure is expected to be biaxial, the limited geometry within the cryostat reduces the ability to sample the anisotropy of the sample. Furthermore, with 4 domain variants expected and a large probe spot size of 3x5 mm, the observed optical response is expected to sample all domain variants, with the in-plane index an aggregate of $n_1$ and $n_2$ and therefore effectively uniaxial. Oscillator fit parameters were fit sequentially, with each parameter initially constrained to within 10% of the room temperature value, and the constraints increasingly expanded until a minimum goodness of fit was achieved. The low temperature fit parameters for the film are given in **Table S2** below. The full collection of raw Ψ, Δ data and model fits for room temperature and low temperature experiments are provided in **Figure S11** below.

**Table S2: Fit parameters for optical constants of BaTiO₃ film at 5 K**

| Oscillator Number | Lorentz | | | Gaussian | | |
| --- | --- | --- | --- | --- | --- | --- |
| | $Amp_i$ | $Br_i$ | $En_i$ | $Amp_i$ | $Br_i$ | $En_i$ |
| 1 | 0.623944 | 3.2068 | 2.001 | 1.733072 | 4.6560 | 3.333 |
| 2 | 6.696964 | 0.9849 | 3.608 | 5.422329 | 0.8694 | 3.589 |
| 3 | 1.693879 | 0.2187 | 4.119 | 1.906398 | 0.2528 | 4.131 |
| 4 | 3.216761 | 0.6612 | 4.627 | 3.182036 | 0.5070 | 4.581 |

Additional fit parameters include:

For the Lorentz oscillator model: $t = 36.53$ nm, $\varepsilon_\infty = 0.708$, $Amp_{IR} = 0.3066$, $En_{IR} = 0\ eV$, $Amp_{UV} = 210.3801$, $En_{UV} = 11.792\ eV$, $d\varepsilon_A = -0.441993$, $d\varepsilon_B = -0.423672$, $d\varepsilon_C = 0.078737$, $d\varepsilon_D = -0.004368$, $d\varepsilon_{IR} = 0.265581$.

For the Gaussian oscillator model: $t = 36.53$ nm, $\varepsilon_\infty = 0.698$, $Amp_{IR} = 0.2730$, $En_{IR} = 0\ eV$, $Amp_{UV} = 210.3801$, $En_{UV} = 11.792\ eV$, $d\varepsilon_A = -0.100343$, $d\varepsilon_B = 0.323656$, $d\varepsilon_C = -0.053430$, $d\varepsilon_D = 0.00089873$, $d\varepsilon_{IR} = 0.219749$.

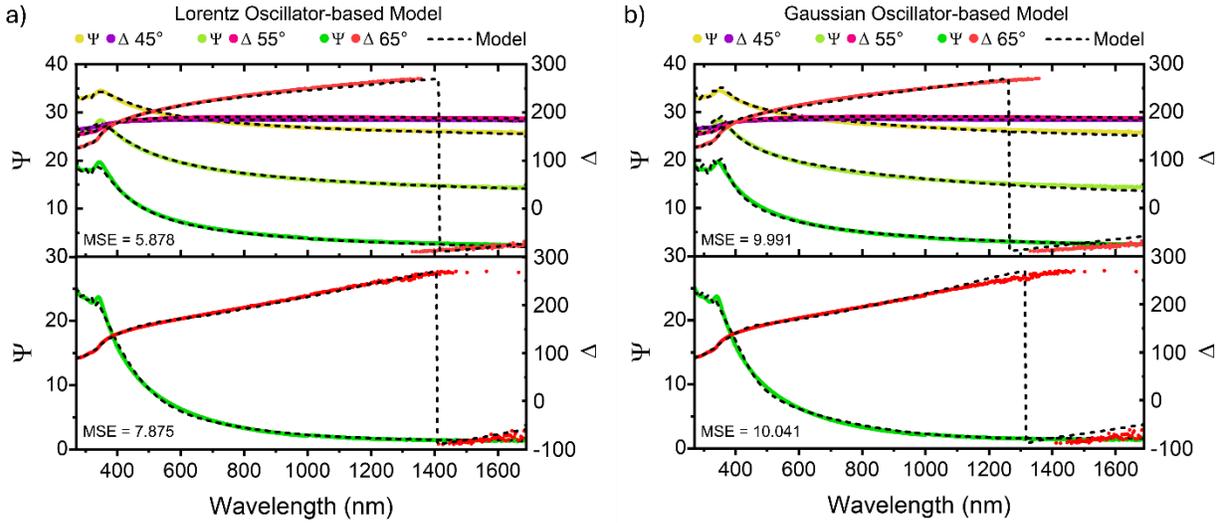

**Figure S11.** Raw $\Psi$, $\Delta$ experimental data collected at room temperature (top) and 5 K (bottom) and resulting fits to data generated using a (a) Lorentz oscillator-based model and (b) Gaussian oscillator-based model as described in Supplemental Note 3. Mean squared error (MSE) values demonstrating the goodness of fit of the models are provided within each plot.

In order to supplement the spectroscopic ellipsometry measurements, an Agilent - Cary 5000 system was used to collect normal incidence UV-Vis spectroscopy specular transmittance measurements on the film that was the focus of the study and similar samples, with the results presented below in **Figure S12**. In addition, a free space near-normal incidence reflectance of the $BaTiO_3$ thin film at 1550 nm was also measured at room temperature. The backside of the substrate was polished at an angle, leading to a wedge geometry that isolates the reflection from only the air-thin film and thin film-substrate interfaces. Collected at a 4° angle of incidence, an absolute normalized reflected intensity of 0.124 was measured. Using a thin film reflection model accounting for multiple reflections and sample birefringence, a predicted value of 0.1237 is obtained using optical constants retrieved from the Gaussian oscillator-based spectroscopic ellipsometry model, indicating excellent agreement.[10] In contrast, the Lorentz oscillator-based model predicts a reflectance of 0.1292, yielding agreement only within 4%. As a consequence of these results, the optical constants provided by Gaussian oscillator-based model are used for evaluation of electro-optic coefficients derived within the main text, as shown in **Figure 2(e-f)**. The dispersion curves obtained by using the Lorentz oscillator-based model are provided below in **Figure S13**.

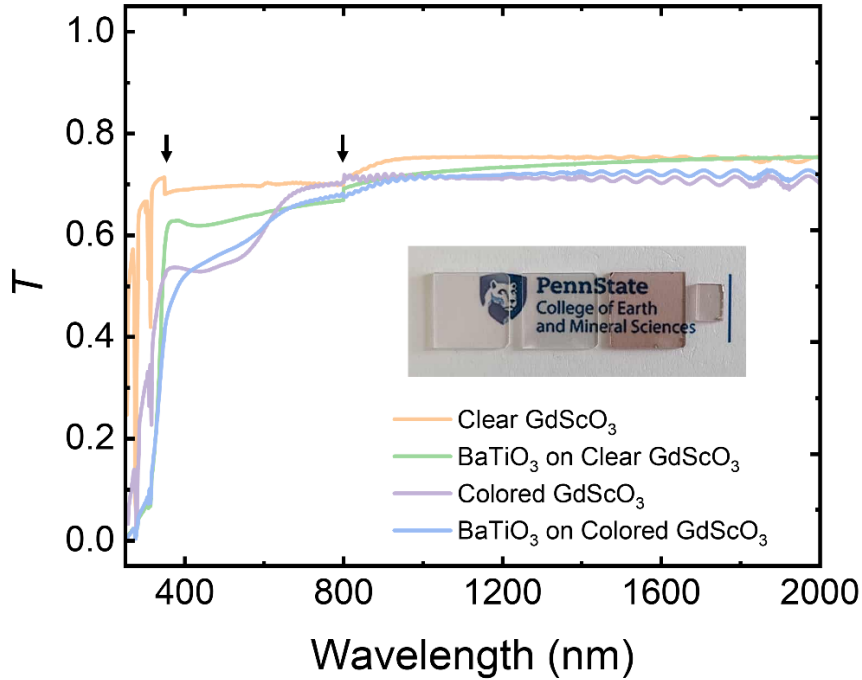

**Figure S12.** Normal incidence transmittance spectra obtained via performing UV-Vis spectroscopy on BaTiO₃ on GdScO₃ film characterized in main text (deposited on colored substrate) along with transmittance spectra collected from a similarly grown film on a visibly transparent GdScO₃ substrate, a transparent GdScO₃ substrate, and a colored GdScO₃ substrate. Variations in the transparency of GdScO₃ substrates can occur due to excess Gd, leading to an orange tint and the reduced transmittance observed in the spectra of the film deposited on a colored substrate and a colored substrate. The discontinuities in the spectra indicated by arrows near 350 nm and 800 nm are due to a change in the light source and detector respectively. (inset) A picture taken showing the optical transparency of the set of samples. From left to right: clear GdScO₃, BaTiO₃ film on clear GdScO₃, colored GdScO₃, BaTiO₃ film on colored GdScO₃. Use of the Penn State mark is granted for this one-time editorial use with no promotional use granted.

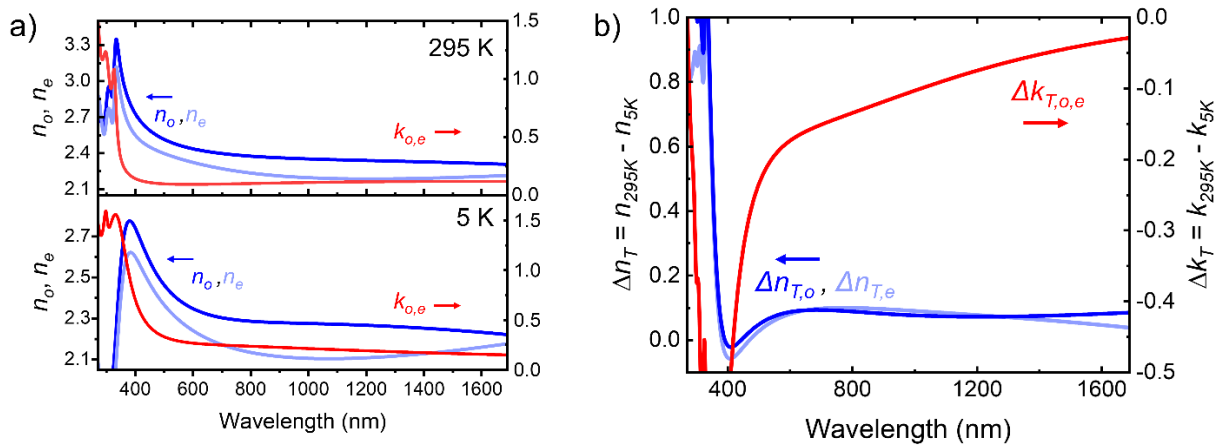

**Figure S13.** a) Refractive index and extinction coefficient of strained BaTiO$_3$ film versus wavelength, at room temperature and at 5 K, as determined by variable angle ellipsometry using a Lorentz oscillator-based model as compared to the Gaussian oscillator-based model presented in **Figure 2(e-f)**. b) Relative refractive index and extinction coefficient changes between room temperature and 5 K, $\Delta n_T = n_{295K} - n_{5K}$, $\Delta k = k_{295K} - k_{5K}$.

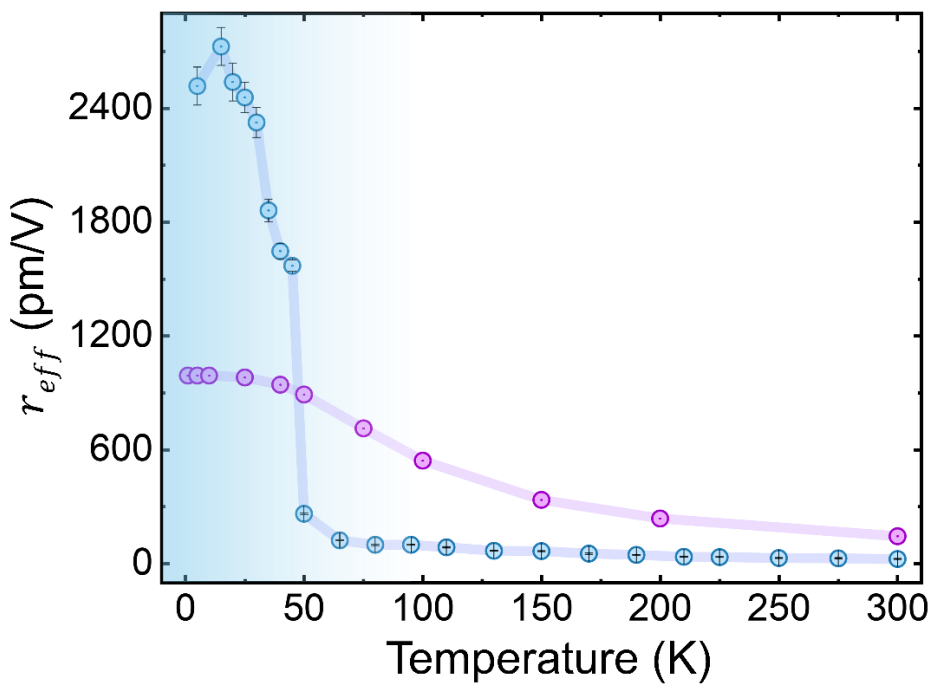

**Figure S14.** Temperature dependence of maximum electro-optic response, derived using a linear interpolated of refractive index values between the experimentally measured values at 295 K and 5 K.

## Note S4: Commentary on the Tensor Nature of the Electro-Optic Response and Behavior of the Index Ellipsoid

In order to optimize the sample geometry for measurement, the tensor nature of the electro-optic response must be diligently considered. The tetragonal BaTiO$_3$ films that are the focus of this work were grown with the [001] direction out of plane so that an in-plane applied electric field would be along the [100] direction, corresponding to the crystal physics direction 3 and 1 respectively. Consequently, the refractive index of the film is described through the perturbed index ellipsoid:[11]

$$\frac{x^2}{n_1^2} + \frac{y^2}{n_2^2} + \frac{z^2}{n_3^2} + 2xz\, r_{51} E_1 = 1. \tag{S1}$$

Note that the cross section obtained by setting $z = 0$ will remain the equation of a circle for a uniaxial material with $n_1 = n_2$ regardless of the magnitude of $E_1$. This means that for a beam propagating along 3, such that the set of polarization states spans the plane spanned by 1 and 2 and the range of corresponding refractive indices experienced by such waves is described by the circle $\frac{x^2}{n_o^2} + \frac{y^2}{n_o^2} = 1$, an applied electric field will not induce an observable refractive index change. For this reason, the film is rotated by 45° in the path of the probe as illustrated by **Figure 2a.**

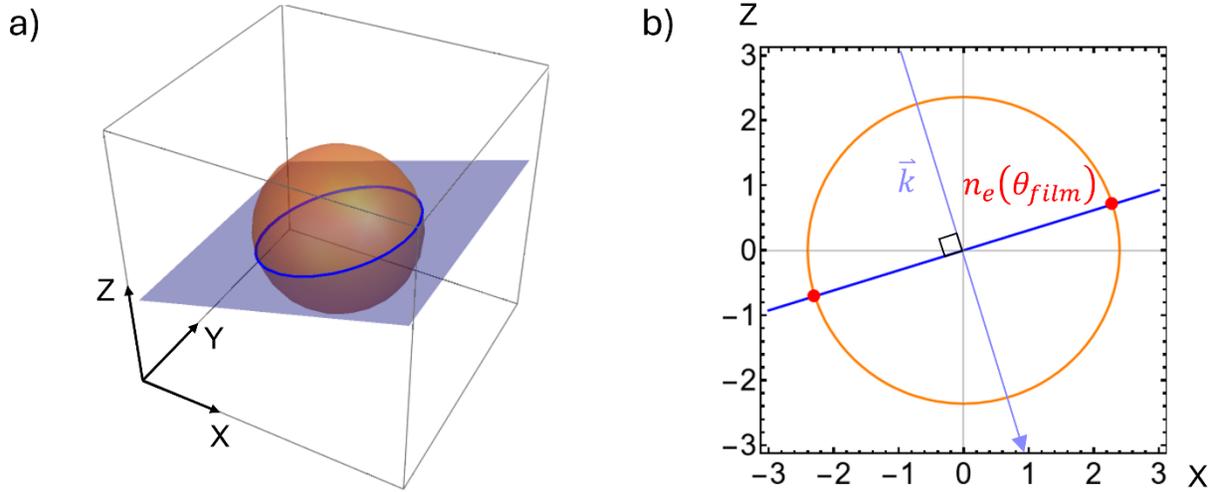

**Figure S15.** a) The intersection of the index ellipsoid described by Equation S1 with $E = 0$ and the plane described by Equation S2(c). The ellipse that defines the intersection carries a semi-minor axis of $n_o$ and a semi-major axis of $n_e(\theta_{film})$. b) The same intersection viewed in the $y = 0$ plane, used to calculate $n_e(\theta_{film})$ as indicated by the red points. A relationship between $n_e(\theta_{film})$ and an applied field $E$ can be used to convert an observed index change to a specific tensor quantity.

Successful interpretation of the electro-optic response relies upon deliberate consideration of the tensor nature of the property, especially as the symmetry of the structure is reduced and more tensor elements contribute to the experimentally measured response. The electro-optic response in the tetragonal phase was previously discussed in main text Section 3.1 and above. To briefly reiterate, the sample geometry was such that only $r_{51}$ contributes to the observed refractive index change. The action of $r_{51}$ results in a nonzero dielectric tensor element in the unperturbed crystal physics coordinate frame, effectively rotating the crystal physics axis about the 2 axis while "stretching" the index ellipsoid in the 1-3 plane. In order to convert the measured $r_{eff}$ to the tensor element $r_{51}$, we geometrically solve for a fixed point on the surface of the perturbed index ellipsoid in accordance with the following prescription.

Consider the ellipsoid in the unperturbed crystal physics axis coordinate system, following **Equation 2** with $E = 0$. The z-axis coincides with the 3 direction and the out-of-plane direction of the sample. An incident probe at 45° propagates through the film at a reduced angle, $\theta_{film}$ according to Snell's law. For simplicity, the splitting of the beam into extraordinary and ordinary rays is not considered, only the propagating angle for the ordinary ray which is roughly 17° from the normal for the unperturbed ellipsoid. With the propagation direction of the probe through the film defined by this angle, one can formulate the equation of a plane describing the span of polarization states for this ray:

$$\vec{p} = \{\cos\theta_{film}, 0, \sin\theta_{film}\} \quad \vec{s} = \{0,1,0\} \tag{S2a}$$

$$\vec{n} = \vec{p} \times \vec{s} \tag{S2b}$$

$$n_x x + n_y y + n_z z = 0 \tag{S2c}$$

The intersection of **Equation S2c** with the index ellipsoid of **Equation 2** yields the cross-sectional ellipse whose perimeter describes the refractive indices experienced by the propagating ray. Since perturbation of the ellipsoid with a field along 1 will leave the 2-axis undisturbed ($2 = 2'$, $n_2 = n'_2 = n_o$), we set $y = 0$ to view the cross-sectional perimeter in the $xz$ plane and the plane of **Equation S2c** as a line. The intersection of these two reveals $n_e(\theta_{film})$. The experimentally measured $\Delta n = n_e(\theta_{film})' - n_e(\theta_{film})$, where $n_e(\theta_{film})'$ is obtained by following the same procedure in the same coordinate system but using the equation for the perturbed ellipsoid under an applied field with $E \neq 0$. With this approach, one can work backwards to relate the measured $\Delta n$ to an exact value of $r_{51}$, as shown in **Figure S12**.

For the monoclinic phase, the general approach remains valid. The crystal physics coordinate system remains unchanged through the phase transformation; the only additional consideration comes from the additional tensor elements present for the monoclinic point group *m*. For an electric field applied along 1, and the monoclinic mirror plane perpendicular to 2, the relevant additional components are the $r_{11}$ and $r_{31}$ coefficients. If we consider a full 4-domain monoclinic structure as suggested by SHG fitting models, then for some domain variants the applied electric field will be along the 2 direction, resulting in contributions from $r_{42}$ and $r_{62}$. For such low symmetry structures, it becomes difficult then to isolate the contribution from a single tensor element.

In order to gauge the response from such other coefficients, EO measurements below 50 K were performed at normal incidence. **Such measurements yielded no detectable response**. Based on the noise floor of the experimental instrumentation, the smallest detectable electro-optic response for a film with the thickness and refractive index of the strained BaTiO$_3$ would correspond to a $r_{eff}$ of 5 pm/V. Utilizing the 4-domain monoclinic model suggested by SHG analysis, we can expect two set of domain twins, referred to as $D_1$ (monoclinic mirror plane perpendicular to tetragonal [100] = 1) and $D_2$ (monoclinic mirror plane perpendicular to tetragonal [010 = 2]). Expressed in the reference frame of the crystal physics axis of the tetragonal system, their respective electro-optic property tensors are:

$$D_1: \begin{pmatrix} 0 & r_{112} & r_{113} \\ 0 & r_{222} & r_{223} \\ 0 & r_{332} & r_{333} \\ 0 & r_{232} & r_{233} \\ r_{131} & 0 & 0 \\ r_{121} & 0 & 0 \end{pmatrix} \quad D_2: \begin{pmatrix} r_{111} & 0 & r_{113} \\ r_{221} & 0 & r_{223} \\ r_{331} & 0 & r_{333} \\ 0 & r_{232} & 0 \\ r_{131} & 0 & r_{133} \\ 0 & r_{122} & 0 \end{pmatrix}$$

Thus, the coefficients that would contribute to a response for a field applied along 1 include $r_{111} = r_{11}$, $r_{221} = r_{21}$, $r_{331} = r_{31}$, $r_{131} = r_{51}$, and $r_{121} = r_{61}$. Under the normal incidence probe with a beam propagating along 3, the effects of $r_{31}$ and $r_{51}$ will not be observed, leaving $r_{11}$, $r_{21}$, and $r_{61}$. The inability to observe an electro-optic response under normal incidence suggests that either the magnitude of all these coefficients in the monoclinic phase is below our 5 pm/V limit, or their magnitudes are comparable such that even though the refractive indices along the 1 or 2 directions change, the birefringence of the material does not. With this in mind we assume that their effects do not contribute to the oblique incidence $r_{eff}$ observed in the monoclinic phase. This leaves only the $r_{31}$ and $r_{51}$ components as possible contributors to the monoclinic response at oblique incidence. From phase-field simulations, the predicted ratio between $r_{51}$ and $r_{31}$ reaches a minimum value of 170:1 at 1 K, as shown in **Figure S16**. Given the large discrepancy in their predicted magnitude, we elect to not consider the contribution from $r_{31}$. Thus, we attribute the measured $r_{eff}$ of the monoclinic response entirely to $r_{51}$, as shown in **Figure 1c** in the main text.

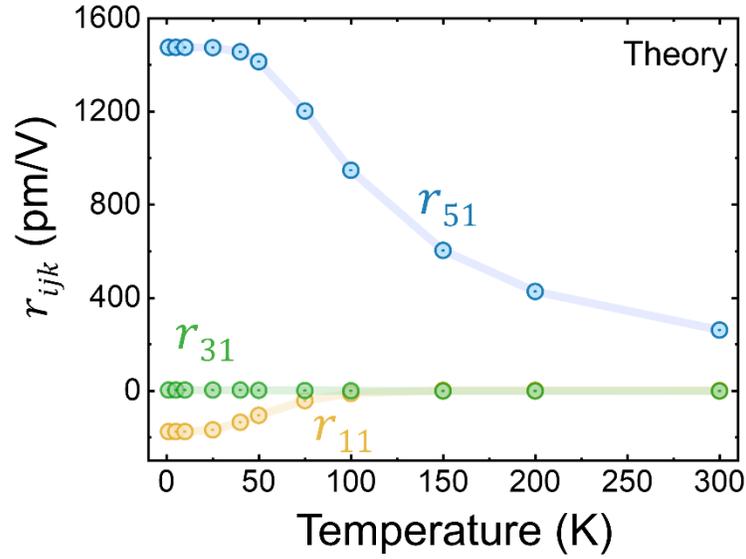

**Figure S16.** Individual electro-optic tensor elements as a function of temperature as predicted by phase-field simulations.

## Note S5: Consistency of Nonlinear Electro-Optic Response with Thermal and Electrical Cycling

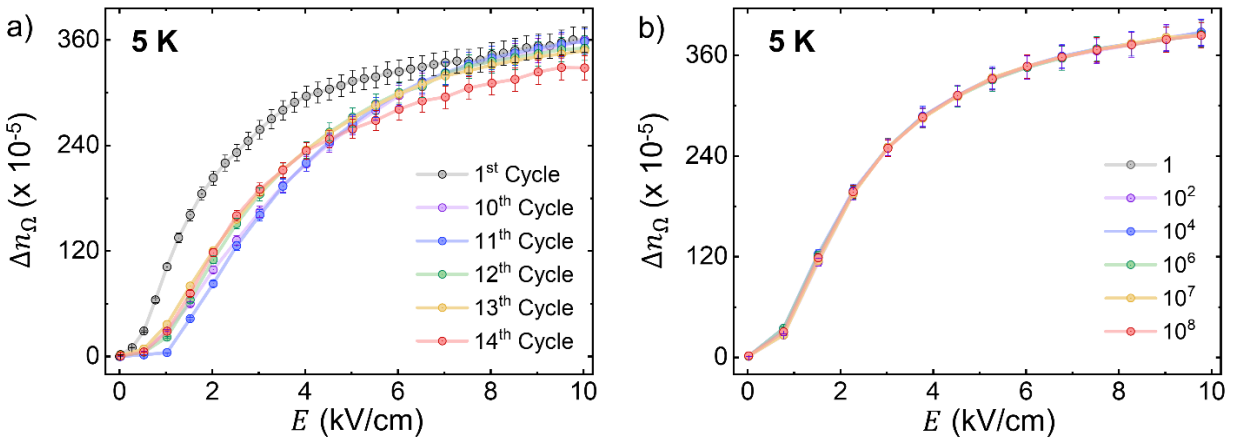

**Figure S17.** a) The enhanced nonlinear electro-optic response observed at 5 K across six different sample cooling runs, indicated with curves of different colors. b) The nonlinear electro-

optic response recorded after subjecting the sample to increasing numbers of cycling under a 5 kV/cm, 20 kHz electric field.

To investigate the reproducibility of the electro-optic response, $\Delta n$ vs $E$ curves were collected and compared across multiple cooling-heating cycles and as a function of electrical cycling, as shown in **Figure S17**. The nonlinearity of the material response was observed to change after every instance the sample was cooled, suggesting a dependence on the initial low-temperature domain structure. The initial cooling cycle of the sample, which yielded the electro-optic data included in **Figures 1, 2,** and **3** of the main text is markedly different from those collected in later experiments. While the speed at which the change in refractive index is reached grew longer with subsequent cycles, the saturated value for the refractive index change remained consistent. These changes can be captured by fitting $r_{(1),max}$, $r_{(1),hf}$, and $r_{(1),lf}$, as described in the main text, with the results shown in **Table S3** below. We note that a reduced density of data points collected in the low field regime in the later cycles resulted in larger uncertainty in the coefficients retrieved from that regime.

**Table S3: Electro-optic coefficients extracted across multiple cooling cycles**

|  | $r_{(1),max}$ (pm/V) | $r_{(1),hf}$ (pm/V) | $r_{(1),lf}$ (pm/V) |
|---|---|---|---|
| Thermal Cycle 1 | 2516 ± 90 | 132 ± 4 | 475 ± 79 |
| Thermal Cycle 10 | 1252 ± 22 | 224 ± 7 | 70 ± 6 |
| Thermal Cycle 11 | 1460 ± 44 | 210 ± 16 | 152 ± 100 |
| Thermal Cycle 12 | 1591 ± 35 | 172 ± 13 | 93 ± 22 |
| Thermal Cycle 13 | 1516 ± 36 | 173 ± 10 | 182 ± 49 |
| Thermal Cycle 14 | 1619 ± 36 | 205 ± 19 | 63 ± 26 |

In comparison to the variation in material response observed with thermal cycling through the ~50 K phase transition, electrical cycling appears to have minimal effect on the nonlinearity of the sample response up to the number of high voltage cycles the sample was subjected to ($10^8$). That is to say, once the system has been cooled below a phase transition the electro-optic response is highly robust and reproducible, but events that allow for the reconfiguration of the low-temperature domains can induce fluctuations in the sample response.

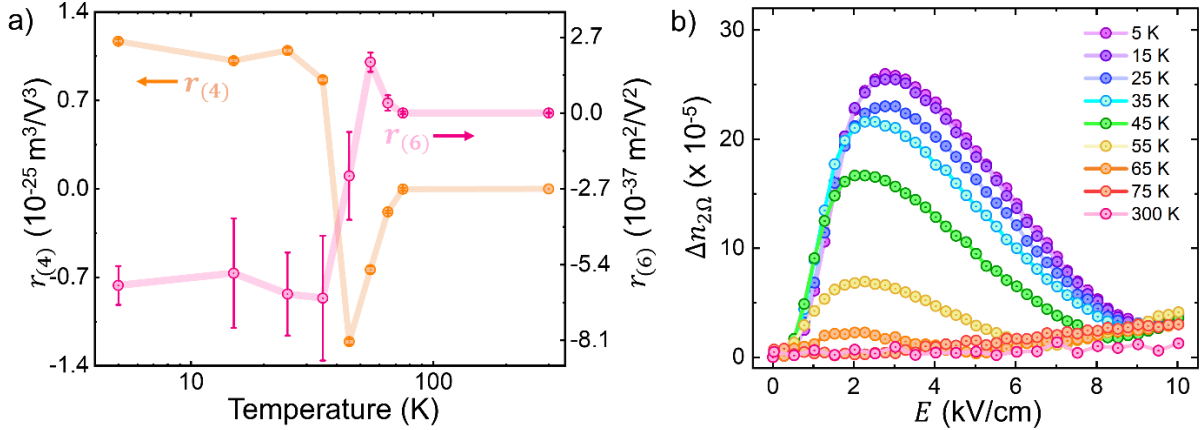

**Figure S18. a)** Fourth ($r_4$) and sixth ($r_6$) order electro-optic coefficients extracted from the low-field regime of the electro-optic response recorded at double the modulation frequency, as discussed in Section 3.2 and shown in **Figure 3c**. **b)** The temperature dependence of the second harmonic electro-optic response, containing even-order higher order electro-optic coefficients.

# Note S6: Phenomenological Model for Fitting the Nonlinear Electro-Optic Response

To capture all three regimes of the nonlinear electro-optic response observed at the modulation frequency with a single equation, an approximate phenomenological model based on the Avrami-like model for the monoclinic phase is proposed:[12]

$$\Delta n = A(1 - \exp(kE^m)) + BE. \qquad (4)$$

The Avrami model is applicable to understanding solid-state transformation kinetics, where the fraction of a transformed phase is expressed as a function of time. The constants $k$ and $m$ are time-independent constants, where $m = D + 1$ reveals the dimensions of growth, $D$, along which the transformation proceeds. In order to capture the full behavior of the S-curve, we add a leading constant $A$ to describe the overall magnitude of the index change and a linear term $BE_{amp}$ to describe the linear response in the saturation regime. We constrain $m$ to be 1 or 2 depending on whether the low-field quadratic regime is present or not (roughly below or above 35 K), leaving $k$ as an independent parameter related to the slope about the inflection point. In the low-field limit, this equation reduces to $\Delta n \sim AkE^2 + BE$ when $m = 2$ and $\Delta n = (Ak + B)E$ when $m = 1$, allowing for the low-field quadratic behavior to be captured if present. In the high-field limit, the equation reduces to $\Delta n = A + BE$. Representative fits of the response at 5 K and 40 K are shown in **Figure S16**.

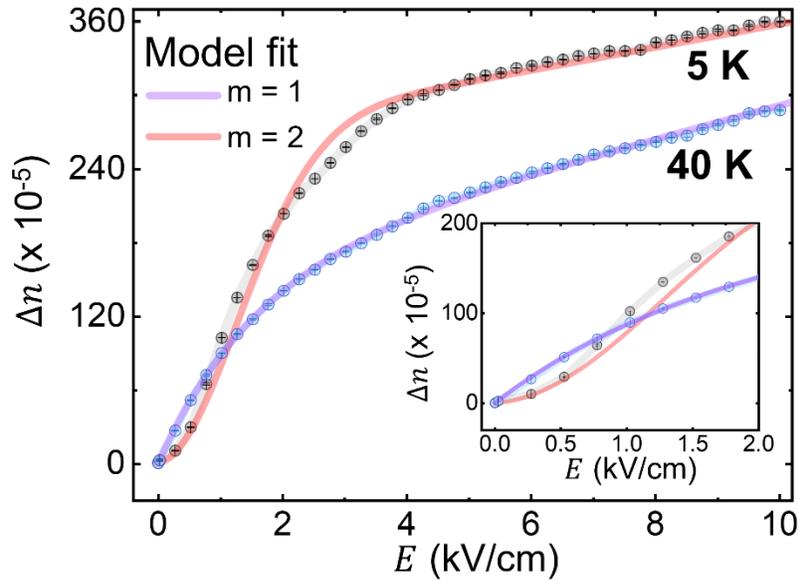

**Figure S19.** Phenomenological model fit of nonlinear response for two representative temperatures, demonstrating the emergence of low-field quadratic behavior.

## Note S7: Second-Harmonic Generation Polarimetry

Due to similarities in crystal structure and symmetry, the model used to fit SHG polarimetry data follows in the framework developed in Ref. [13].[13] For convenience, much of the information is reproduced below at the author's permission:

A schematic of the SHG setup is shown in Supplemental **Figure S17**. The lab coordinates (X, Y, Z) are attached to the direction of the incoming beam: X ∥ *p* polarization, Y ∥ *s* polarization, where *p* and *s* polarization states are set by a rotating half wave plate. The crystal axes coordinate *i* = (1,2,3) are attached to the sample and related to the substrate as follows:

For GdScO$_3$ substrates: 1 ∥ [001]$_o$ , 2 ∥ [1-10]$_o$ , 3 ∥ [110]$_o$ , where subscript "o" refers to orthorhombic unit cell of the scandate substrates.

In lab coordinates (X, Y, Z), the electric field of the incident beam can be written as $(E_o \cos \varphi, E_o \sin \varphi, 0)$ where $\varphi$ is the polarization rotation angle introduced by the half wave plate as shown in **Figure S20**. For an incidence angle θ on the sample, the electric field in the crystal axes coordinates can be expressed as $(E_o \cos \varphi \cos \theta, E_o \sin \varphi, -E_o \cos \varphi \sin \varphi)$. The induced nonlinear polarization $P^{2\omega}$, is related to the incident electric field through the nonlinear susceptibility tensor, $d_{ijk}$ through the following equation:

$$P_i^{2\omega} \propto d_{ijk} E_j^{\omega} E_k^{\omega} \tag{1}$$

The proportionality constants depend on incident beam fluence, Fresnel's coefficients at the film-air and film-substrate interfaces and the thickness of the films.

**Tetragonal model**:

For the tetragonal phase of point group 4*mm*, the nonlinear susceptibility can be written in Voigt notation as:

$$d_{ij} = \begin{pmatrix} 0 & 0 & 0 & 0 & d_{15} & 0 \\ 0 & 0 & 0 & d_{15} & 0 & 0 \\ d_{31} & d_{31} & d_{33} & 0 & 0 & 0 \end{pmatrix}$$

The induced nonlinear polarization in the crystal coordinates *i* = (1,2,3) (calculated through equation 1) can be rotated back to lab coordinates (X, Y, Z), to give the *p* and *s*-polarized components of the SHG (*p* ∥ X and *s* ∥ Y). The *p* and *s*-polarized SHG intensities can be expressed as follows:

<u>For normal incidence ($\theta = 0°$)</u>

$$I_p^{2\omega} \propto \left(P_p^{2\omega}\right)^2 = 0$$

$$I_s^{2\omega} \propto (P_s^{2\omega})^2 = 0$$

For oblique incidence ($\theta = 45°$)

$$I_p^{2\omega} \propto \left(P_p^{2\omega}\right)^2 \propto ((2d_{15} - d_{31} - d_{33})\cos[\varphi]^2 - 2d_{31}\sin[\varphi]^2)^2 \tag{2}$$

$$I_s^{2\omega} \propto (P_s^{2\omega})^2 \propto d_{15}^2 \sin[2\varphi]^2$$

**Monoclinic model**:

For the low temperature monoclinic phase with point group $m$, a multi-domain model is assumed. In crystal physics coordinates $i = (1,2,3)$, nonlinear susceptibility in the monoclinic phase can be written as:

$$d_{ij} = \begin{pmatrix} d_{11} & d_{12} & d_{13} & 0 & d_{15} & 0 \\ 0 & 0 & 0 & d_{24} & 0 & d_{26} \\ d_{31} & d_{32} & d_{33} & 0 & d_{35} & 0 \end{pmatrix}$$

Here, the monoclinic mirror plane is perpendicular to the crystallographic $b$-axis of the monoclinic cell of $BaTiO_3$ which is parallel to $i = 2$ crystal physics coordinate. Through rotating the unit cell by right angles, 4 such unit cells can be achieved, each associated with an area fraction labelled as below:

Domain 1: $a \parallel 1, b \parallel 2, c \parallel 3$ (Area fraction: $A_1$)
Domain 2: $a \parallel 2, b \parallel -1, c \parallel 3$ (Area fraction: $A_2$)
Domain 3: $a \parallel -1, b \parallel -2, c \parallel 3$ (Area fraction: $A_3$)
Domain 4: $a \parallel -2, b \parallel 1, c \parallel 3$ (Area fraction: $A_4$)

Here ($a$, $b$, $c$) denote the crystallographic axes of the monoclinic unit cell, $i = (1,2,3)$ denote the previously defined crystal physics coordinate system and the area fractions are constrained to $A_1 + A_2 + A_3 + A_4 = 1$.

For normal incidence ($\theta = 0°$)

From Equation 1, the induced nonlinear polarization can be calculated in the crystal coordinate system and can be transformed into lab coordinates (X, Y, Z). The resultant nonlinear polarization in the lab coordinates X (*p*-polarized) and Y (*s*-polarized) are tabulated below (Table S1):

**Table S4: *p* and *s*-polarization components for different domains**

|  | $P_p^{2\omega}$ | $P_s^{2\omega}$ |
|---|---|---|
| Domain 1 | $d_{11}\cos^2\varphi + d_{12}\sin^2\varphi$ | $d_{26}\sin 2\varphi$ |
| Domain 2 | $d_{26}\sin 2\varphi$ | $d_{12}\cos^2\varphi + d_{11}\sin^2\varphi$ |
| Domain 3 | $-(d_{11}\cos^2\varphi + d_{12}\sin^2\varphi)$ | $-d_{26}\sin 2\varphi$ |
| Domain 4 | $-d_{26}\sin 2\varphi$ | $-(d_{12}\cos^2\varphi + d_{11}\sin^2\varphi)$ |

The effective SHG intensity can be calculated as follows:

$$I_p^{2\omega} \propto \left(P_p^{2\omega}\right)^2 \propto \left(A_1 P_{p,\text{domain 1}}^{2\omega} + A_2 P_{p,\text{domain 2}}^{2\omega} + A_3 P_{p,\text{domain 3}}^{2\omega} + A_4 P_{p,\text{domain 4}}^{2\omega}\right)^2$$

This can be simplified to:

$$I_p^{2\omega} \propto K_{1p}\left(\sin^2\varphi + K_{2p}\cos^2\varphi\right)^2 + K_{3p}\sin^2 2\varphi + K_{4p}\left(\sin^2\varphi + K_{2p}\cos^2\varphi\right)\sin 2\varphi$$

(3)

where,

$$K_{1p} = \delta A_1^2 d_{12}^2 \quad K_{2p} = \frac{d_{11}}{d_{12}} \quad K_{3p} = \delta A_2^2 d_{26}^2 \quad K_{4p} = 2\delta A_1 \delta A_2 d_{12} d_{26}$$

$\delta A_1 = A_1 - A_3$ and $\delta A_2 = A_2 - A_4$. Similar expressions can also be derived for $I_s^{2\omega}$ with the following coefficients:

$$K_{1s} = \delta A_2^2 d_{11}^2 \quad K_{2s} = \frac{d_{12}}{d_{11}} \quad K_{3s} = \delta A_1^2 d_{26}^2 \quad K_{4s} = 2\delta A_1 \delta A_2 d_{11} d_{26}$$

For oblique incidence ($\theta = 45°$):

Equations for oblique incidence polarimetry involve more convoluted combinations of $d$-coefficients and area fractions, however they can be reduced to the same form as equation (3). The exact convolutions are listed below:

$$K_{1p} = \frac{1}{8}[2(A_1 + A_3)d_{32} + 2\delta A_1 d_{12} + 2d_{31}(A_2 + A_4)]^2$$

$$K_{2p} = \frac{1}{2\sqrt{2K_{1p}}}[\delta A_1(d_{11} + d_{13} - 2d_{35}) + (A_1 + A_3)(d_{31} + d_{33} - 2d_{15}) + (A_2 + A_4)(d_{32} + d_{33} - 2d_{24})]$$

$$K_{3p} = \frac{[\delta A_2(d_{26} - d_{35})]^2}{4}$$

$$K_{4p} = 2\sqrt{K_{1p}K_{3p}}$$

$$K_{1s} = \frac{1}{4}(A_2 - A_4)^2 d_{11}^2$$

$$K_{2s} = \frac{1}{2\sqrt{K_{1s}}}(d_{12} + d_{13})(A_2 - A_4)$$

$$K_{3s} = \frac{1}{4}\left(\sqrt{2}A_1(d_{26} - d_{24}) - \sqrt{2}A_3(d_{26} + d_{24}) + \frac{d_{15}}{\sqrt{2}}(A_4 - A_2)\right)^2$$

$$K_{4s} = 2\sqrt{K_{1s}K_{3s}}$$

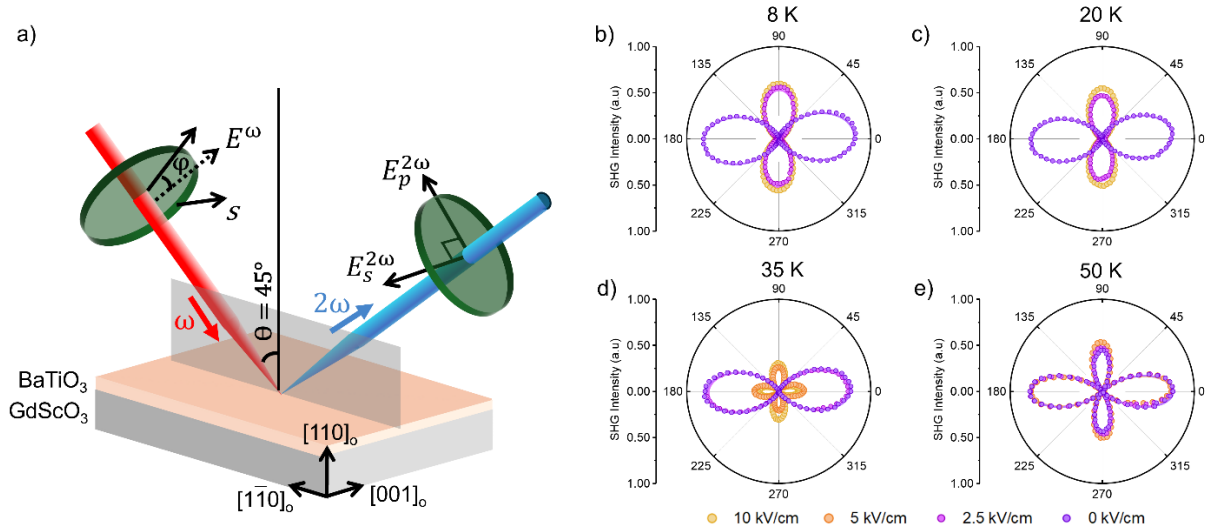

**Figure S20.** a) Schematic of the SHG experimental setup, where a fundamental probe at frequency $\omega$ generates second harmonic light at frequency $2\omega$. Crystallographic orientations of the orthorhombic GdScO$_3$ substrate are indicated. b-e) Applied electric field dependent $I_p^{2\omega}$ polarimetry curves taken between 8 K and 50 K.

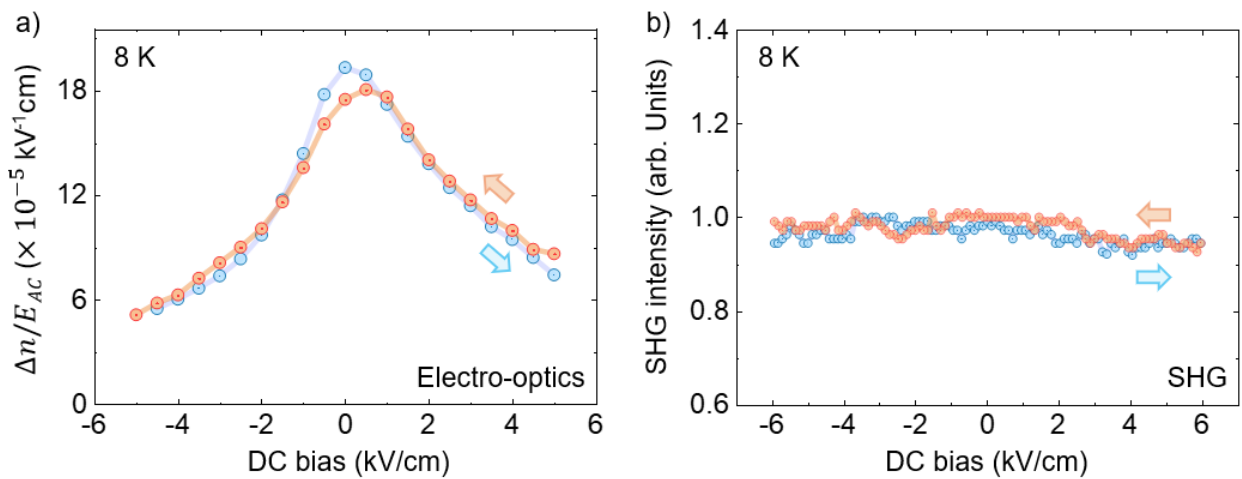

**Figure S21**: a) Hysteretic behavior of the electro-optic response of the BaTiO$_3$/GdScO$_3$ sample at low temperature obtained by sweeping the DC bias in the forward and reverse directions while using an AC electric field to measure the electric field-induced change in refractive index. b) SHG-DC bias loop measured at 8 K showing minimal hysteresis upon field sweep.

**Appendix A: Detailed expression of the free energy function and the associated parameters**

For BaTiO$_3$, we use the cubic phase ($m\bar{3}m$) as our high symmetry reference state and employ an 8th-order landau expansion describe the relative stability of the lattice polarization compared to the cubic reference state. The coefficients used for this paper are adjusted from [14] and [15] to include the effect of cryogenic fluctuations and are given in **Table S5**.

**Table S5.** Coefficients in the thermodynamic free energy function and equation of motion for BaTiO$_3$

| | | | |
|---|---|---|---|
| $g^{LL}_{1111}$ | $18.5 \times 10^{-2}$ (m$^4$/C$^2$) | $p_{1111}$ | 0.5328 (Unitless) |
| $g^{LL}_{1122}$ | $2.5 \times 10^{-2}$ (m$^4$/C$^2$) | $p_{1122}$ | 0.1584 (Unitless) |
| $g^{LL}_{1212}$ | $12.85 \times 10^{-2}$ (m$^4$/C$^2$) | $p_{1212}$ | $-0.432$ (Unitless) |
| $\mu_e$ | $35.5 \times 10^{-23} \left(\frac{Kg\, m^4}{m\, C^2}\right)$ | $\gamma_e$ | $3 \times 10^{-9} \left(\frac{Kg\, m^4}{ms\, C^2}\right)$ |
| $a_{11}$ | $a_0 T_s \left(\coth\left(\frac{T_s}{T}\right) - \coth\left(\frac{T_s}{T_c}\right)\right)$ | $B^{e,ref}_{ij}(T_0)$ | 0.2356 (Unitless) |
| $T_c$ | 388K | $T_0$ | 398K |
| $T_s$ | 54 K | $a_0$ | $4.124 \times 10^5 \left(\frac{J}{m^3}\, \frac{m^4}{K\, C^2}\right)$ |
| $a_{1111}$ | $-2.097 \times 10^8 \left(\frac{J}{m^3}\, \frac{m^8}{C^4}\right)$ | $\alpha_{11}$ | $2.657 \times 10^{-5} \left(\frac{1}{K}\right)$ |
| $a_{1122}$ | $7.974 \times 10^8 \left(\frac{J}{m^3}\, \frac{m^8}{C^4}\right)$ | $a_{11111111}$ | $3.863 \times 10^{10} \left(\frac{J}{m^3}\, \frac{m^{16}}{C^8}\right)$ |
| $a_{111111}$ | $1.294 \times 10^9 \left(\frac{J}{m^3}\, \frac{m^{12}}{C^6}\right)$ | $a_{11111122}$ | $2.529 \times 10^{10} \left(\frac{J}{m^3}\, \frac{m^{16}}{C^8}\right)$ |
| $a_{111122}$ | $-1.95 \times 10^9 \left(\frac{J}{m^3}\, \frac{m^{12}}{C^6}\right)$ | $a_{11112222}$ | $1.637 \times 10^{10} \left(\frac{J}{m^3}\, \frac{m^{16}}{C^8}\right)$ |
| $a_{112233}$ | $-2.509 \times 10^9 \left(\frac{J}{m^3}\, \frac{m^{12}}{C^6}\right)$ | $a_{11112222}$ | $1.637 \times 10^{10} \left(\frac{J}{m^3}\, \frac{m^{16}}{C^8}\right)$ |

| $C_{11}$ | $1.78 \times 10^{11}$ Pa | $a_{11112233}$ | $1.367 \times 10^{10} \left( \frac{J}{m^3} \frac{m^{16}}{C^8} \right)$ |
| --- | --- | --- | --- |
| $C_{12}$ | $0.964 \times 10^{11}$ Pa | $C_{44}$ | $1.22 \times 10^{11}$ Pa |